\documentclass[aip,jap,reprint,superscriptaddress,nofootinbib]{revtex4-1}

\usepackage{amssymb,amsmath,bm,graphicx,xcolor}

\usepackage{lipsum}

\usepackage{subcaption}

\usepackage{epstopdf}

\usepackage{makecell}

\usepackage[symbol]{footmisc}

\begin{document}

\title{Enhanced low-flux sensitivity (ELFS) effect of neutron-induced displacement damage in bipolar devices: physical mechanism and parametric model
}

\author{Yang Liu}
\email{liuyang@mtrc.ac.cn}
\affiliation{Microsystem and Terahertz Research Center, China Academy of Engineering Physics,
Chengdu 610200, P.R. China}
\affiliation{Institute of Electronic Engineering, China Academy of Engineering Physics,
Mianyang 621999, P.R. China}

\author{Ying Zhang}
\affiliation{Microsystem and Terahertz Research Center, China Academy of Engineering Physics,
Chengdu 610200, P.R. China}
\affiliation{Institute of Electronic Engineering, China Academy of Engineering Physics,
Mianyang 621999, P.R. China}

\author{Mu Lan}
\affiliation{Microsystem and Terahertz Research Center, China Academy of Engineering Physics,
Chengdu 610200, P.R. China}
\affiliation{Institute of Electronic Engineering, China Academy of Engineering Physics,
Mianyang 621999, P.R. China}

\author{Chunsheng Jiang}
\affiliation{Microsystem and Terahertz Research Center, China Academy of Engineering Physics,
Chengdu 610200, P.R. China}
\affiliation{Institute of Electronic Engineering, China Academy of Engineering Physics,
Mianyang 621999, P.R. China}

\author{Yu Song}
\email{songyu@mtrc.ac.cn}
\affiliation{Microsystem and Terahertz Research Center, China Academy of Engineering Physics,
Chengdu 610200, P.R. China}
\affiliation{Institute of Electronic Engineering, China Academy of Engineering Physics,
Mianyang 621999, P.R. China}
\date{\today}

\begin{abstract}
Similar to the enhanced low-dose-rate sensitivity (ELDRS) effect of ionization damage,
an enhanced low-flux senstivity (ELFS) effect has been reported in
ions/neutron irradiation on n-type silicon or PNP transistors.
However, the existing mechanism and simulation dominated by the diffusion dynamics
give much higher transition flux than the experimental obseverations.
In this work, we develop a new model based on the annealing of defect clusters for the ELFS effect.
Simulations considering
Si-interstitial-mediated inter-cluster interactions during their annealing processes
successfully reproduce the ELFS effect.
The ratio of Si interstitials captured by defect clusters to those dissipating off on the sample edges or re-merging into the bulk
is found as the key parameter dominating the enhancement factor (EF) of the ELFS effect. 
We also establish a compact parametric model based on the mechanism,
which  is found to 
provide a good quantitative description of the experimental results.
The model predicts the existence of nonsensitive regions at sufficiently low and high fluxes
as well as a non-trivial fluence and temperature dependence of the enhancement factor.
\end{abstract}

\maketitle

\section{Introduction}
\label{sec:Introduction}

Displacement damage (DD) in irradiated Si bipolar devices causes the degradation of electronic properties of the systems.
In most operational amplifiers and comparators,
the degradation of the input bias current ($I_{iB}$) is usually
the most sensitive parameters. 
The most sensitive components with respect to the degradation of $I_{iB}$ are the input-stage bipolar junction transistors (BJTs)
~\cite{Pease1996, Barnaby1999, Barnaby2001, Barnaby2002}.
Numerous reports~\cite{Srour2003, Srour2013} have studied various DD effects
and the underlying mechanisms are attributed to the formation of several defect structures including
point defects and small clusters\cite{Bracht1998, Cowern1999}, rod like (\{311\}) defects \cite{Li1998, Eaglesham1994}, and
dislocations \cite{Li1998, Boninelli2006}.
These defects change the recombination rate of charge carriers in the base region of the transistors and may also cause donor/acceptor compensation,
which remarkably influence the electronic properties of 
BJTs.
In the experiments of heavy ions implantation with high fluence ($>1\times10^{12}~\text{cm}^{-2}$) and high flux
($>1\times10^{10}~\text{cm}^{-2}\text{s}^{-1}$),
the more densed particle bombardment is found to enhance the DD at higher irradiation flux~\cite{Holland1985, Haynes1991}.
For the situations of lower fluence ($<1\times10^{10}~\text{cm}^{-2}$) and
flux ($1\times10^{7}\text{-}2\times10^{10}~\text{cm}^{-2}\text{s}^{-1}$),
there are also reports saying that the DD can decrease with increasing flux \cite{Svensson1995, Svensson1997}.

Neutron induces the highest NIEL/IEL ratio among the particles, where NIEL and IEL stand for non-ionizing energy loss and ionization energy loss, respectively.
However, the study of the flux dependence of neutron-induced DD 
is rarely seen.
There are suspicions of the presence of the flux effect of neutron iradiations, based on considerations as follows.
1) Different from the heavy ions with MeV energies,
the neutron impacts normally do not generate a prominent long trace of defects;
Instead, the damage clusters are smaller and more localized in space.
2) The average distance between the adjacent impacts for neutron irradiation is huge, which is larger than $1\mu m$
even at a total fluence of $1\times10^{12}~\text{cm}^{-2}$.
The experimental observations show that there is rarely defect clusters with radius larger than $5~nm$
\cite{Donnelly2003, Narayan1981, Jencic1995},
thus the direct overlap among clusters is less possible.
Estimated by Gossick's theory \cite{Gossick1959, Curtis1973},
the maximum range of the influence of the electric potential distorted by a single cluster is hardly larger than $1~\mu m$.
Therefore, it is naturally to see the effects of neutron irradiations as the accumulated events of single-particle incidence,
thus the DD is thought to unlikly show flux effects.
Actually, results supporting that there is no flux effect 
has been reported \cite{zontar1999}.

To clarify the influence of the flux on the neutron induced DD in silicon bipolar devices, 
very recently, we did low equivalent fluence ($1\times10^{10}~\text{cm}^{-2}$) and various flux ($5\times10^5-5\times10^6~\text{cm}^{-2}\text{s}^{-1}$) neutron irradiation experiments on bipolar analogy circuits and BJTs \cite{Zhang2019}.
In contrast to the results of heavy ion bombardment results of high fluence,
our 
experiments have shown
evident enhanced low flux sensitivity (ELFS) effects of the DD in
both integrated circuits and discrete transistors.
To identify the dominate mechanisms,
the characteristic time of different possible processes of DD were estimated
and compared with the experimental conditions.
It is found that,  
the mechanisms responsible for the flux dependence
arise in the processes of short-term annealing, i.e.
the interactions among sequential impacts influence the formation of the stable damage.
Some previous research suggest that the flux dependent results come from the interactions among defect clusters
whose reaction rates are limited by the diffusion of Si interstitials \cite{Hallen1991, Svensson1993, Svensson1997}.
However, there are nonnegligible discrepancy of the transition flux between the calculated results and
the experimental observations.
The transitions flux derived from simulations are much higher than the experimental results
which will be discussed in the main body of this paper.

\begin{figure}[!t]
  \includegraphics[width=0.43\textwidth]{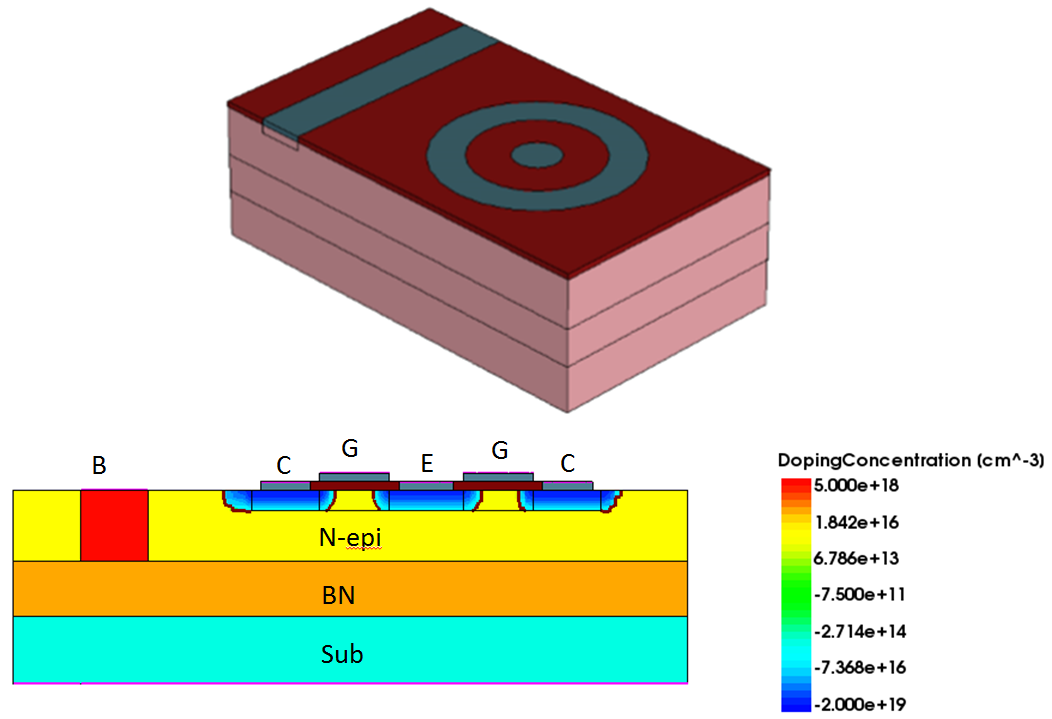}
  \caption{Schematic of the cross section of GLPNP BJT.}
\label{fig:GLPNP_config}
\end{figure}

In this work, we investigate the reasons and propose a defect clusters' self-annealing limited model and a parametric model
for the ELFS effect of neutron-induced DD.
The proposed model considers the cluster nature of the neutron-induced displacement defects and suggest the reactions of
inter clusters are limited by the reordering processes of each cluster.
To test the mechanisms, numerical simulations are performed.
The charged states of defects and
the mobility enhancement effect of mobile particles promoted by exchanging charge carriers with the environment are also considered
\cite{BarYam1984(1),BarYam1984(2),Car1984}.
The simulated results show qualitatively agreements with the experimental observations and
predict an inverse S-shaped flux dependence. 
Further analyses of the mechanisms of the flux effects leads to the construction of an analytical model.
The modeling results show good quantitative agreements with the experimental results and predict
that the DD depends on not only the flux but also on the total fluence and irradiation temperature.

The paper is organized as following.
In Sec. \ref{sec:Experiment}, the experimental results are simply shown.
In Sec. \ref{sec:isolateddefectsmodel}, the conventional model \cite{Hallen1991, Svensson1993, Svensson1997} and simulated results are given.
The discrepancy between the simulated results and the experimental observations are discussed and
the reasons are attributed to the formation of metastable defect complexes.
A modified model is demonstrated in Sec. \ref{sec:clustermodel},
by which the simulated results show better agreement with the experimental results.
In Sec. \ref{chap:mechanisms}, we analysis the mechanisms of the flux effect and show that
a retarded recombination mechanisms during the annealing process of defect clusters dominates the effect.
Based on the mechanisms, a compact parametric model is proposed in Sec.~\ref{sec:Parametricmodel}.
The model gives analytical solutions of DD with respect to the irradiation flux, fluence and temperature.
Some modifications that may need to be made in future are discussed in Sec.~\ref{sec:discussion}.
Conclusions are gathered in Sec.~\ref{sec:conclusion}.

\begin{figure}[!t]
  \includegraphics[width=0.4\textwidth]{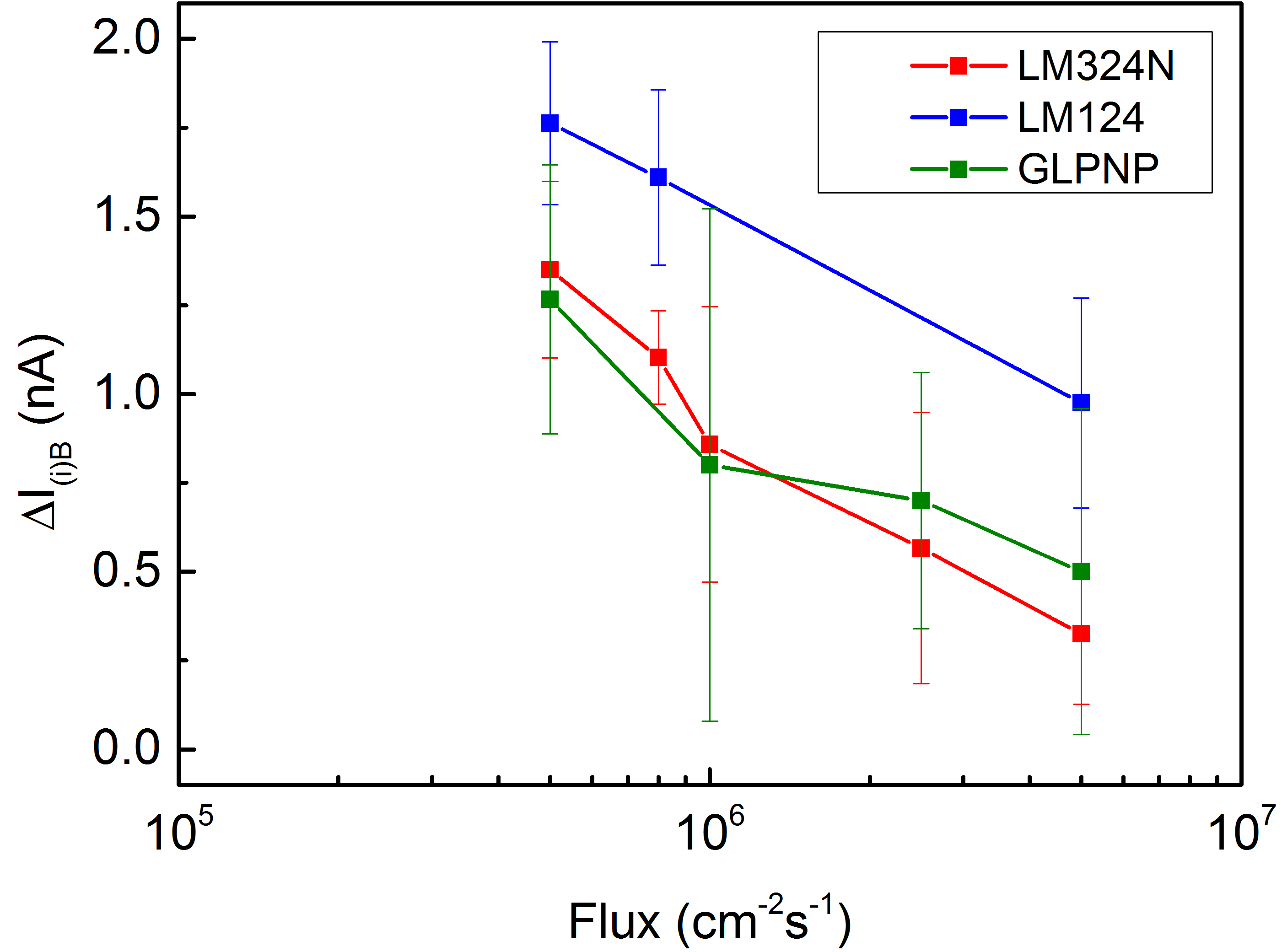}
  \caption{The increase of the input bias current ($\Delta I_{(i)B}$) versus neutron flux for
LM324N, LM124 and GLPNP. All samples are exposed to the same total fluence of $10^{10}~cm^{-2}$.}
\label{fig:Exptresults}
\end{figure}

\section{Experimental results overview}
\label{sec:Experiment}

The detailed characteristics of the experimental results have been analysed in details in one of our previous papers\cite{Zhang2019},
here we just briefly explain the settings of the experiments and
emphasis the key characteristics of the results of the flux effects.

Two kinds of operational amplifier, LM324N 
and LM124 produced by Texas Instruments,
and a kind of  gate lateral PNP (GLPNP) BJTs 
are used as the research objects.
In LM324N and LM124, the input bias currents are most sensitive to the base currents of their input-stage PNP BJTs.
The configuration of the GLPNP BJT is illustrated in Fig.~\ref{fig:GLPNP_config}.
Neutron irradiations were performed at the Chinese Fast Burst Reactor-II
(CFBR-II) of Institute of Nuclear Physics and Chemistry, China Academy of Engineering Physics.
The reactor provides a controlled 1MeV equivalent neutron irradiation.
All samples were irradiated to a total fluence of $1\times10^{10}~\text{cm}^{-2}$ at room temperature,
at flux varying between
$5\times10^{5}~\text{cm}^{-2}\text{s}^{-1}$ and $5\times10^{6}~\text{cm}^{-2}\text{s}^{-1}$.
During the irradiation, all pins are shorted and grounded.

\begin{figure*}[!t]
  \begin{subfigure}[b]{0.22\textwidth}
    \includegraphics[width=\textwidth]{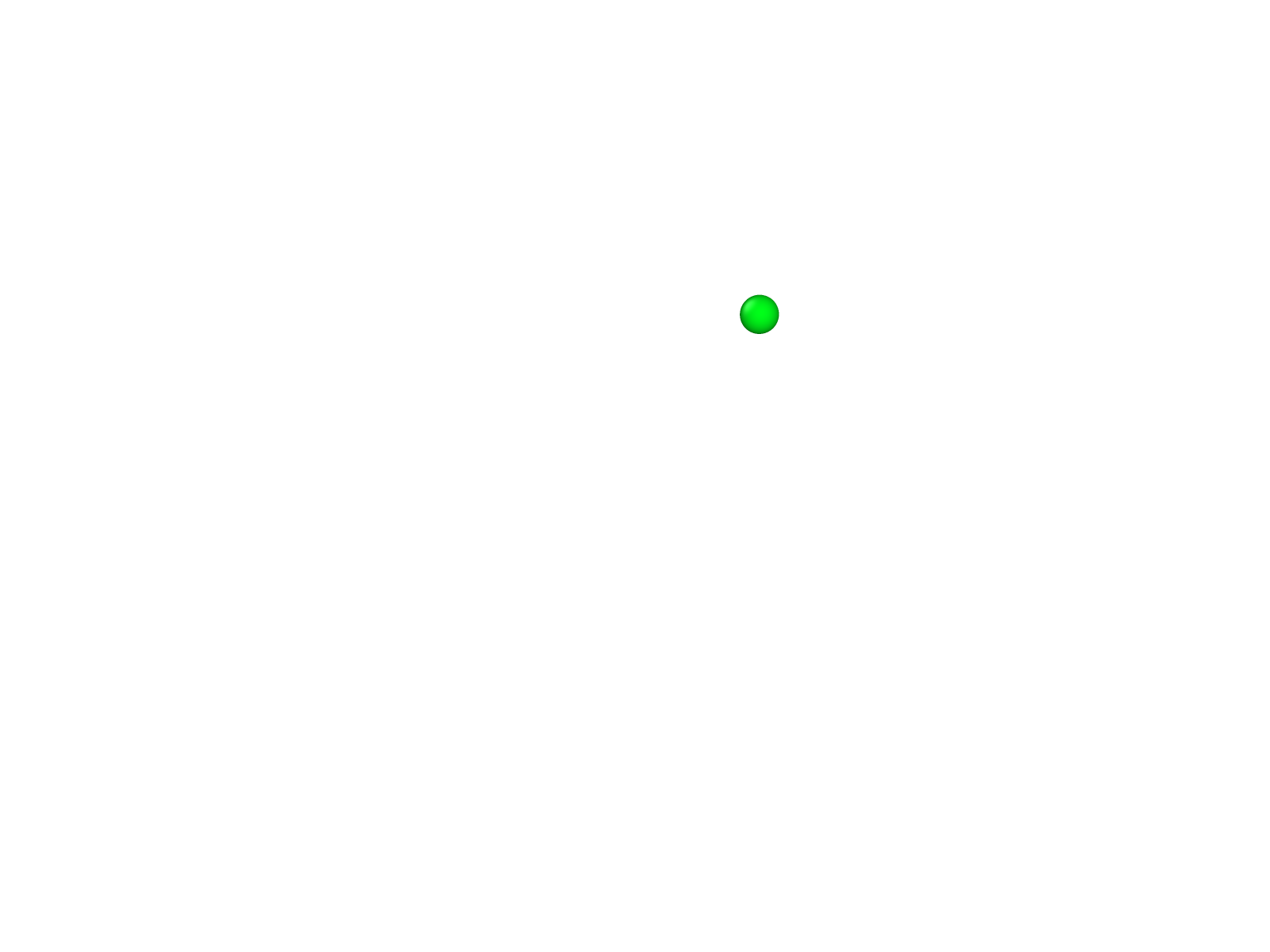}
    \caption{$t=0$}
  \label{fig:LAMMPS0}
  \end{subfigure}
~
  \begin{subfigure}[b]{0.22\textwidth}
    \includegraphics[width=\textwidth]{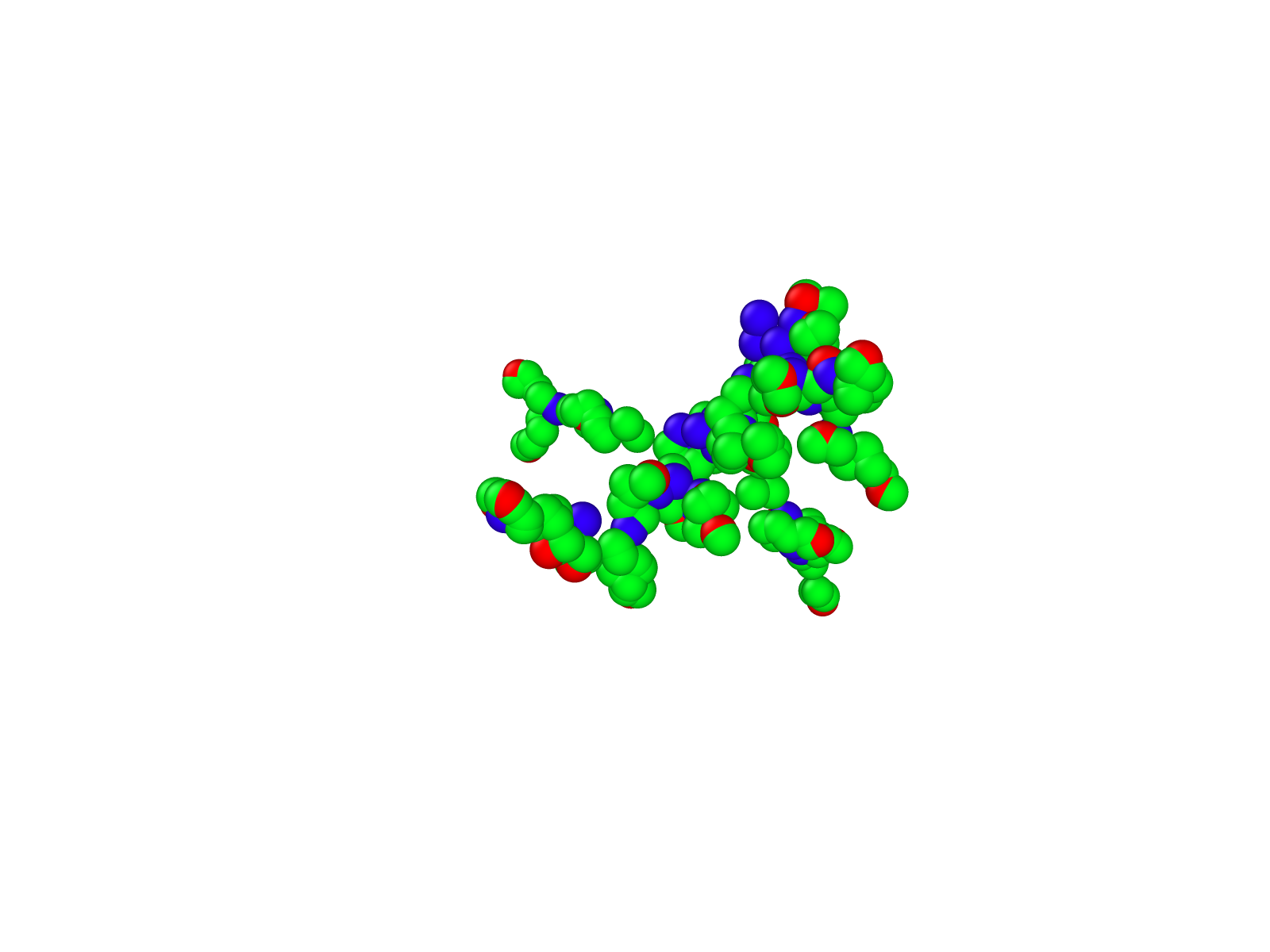}
    \caption{$t=0.1ps$}
  \label{fig:LAMMPS1}
  \end{subfigure}
~
  \begin{subfigure}[b]{0.22\textwidth}
    \includegraphics[width=\textwidth]{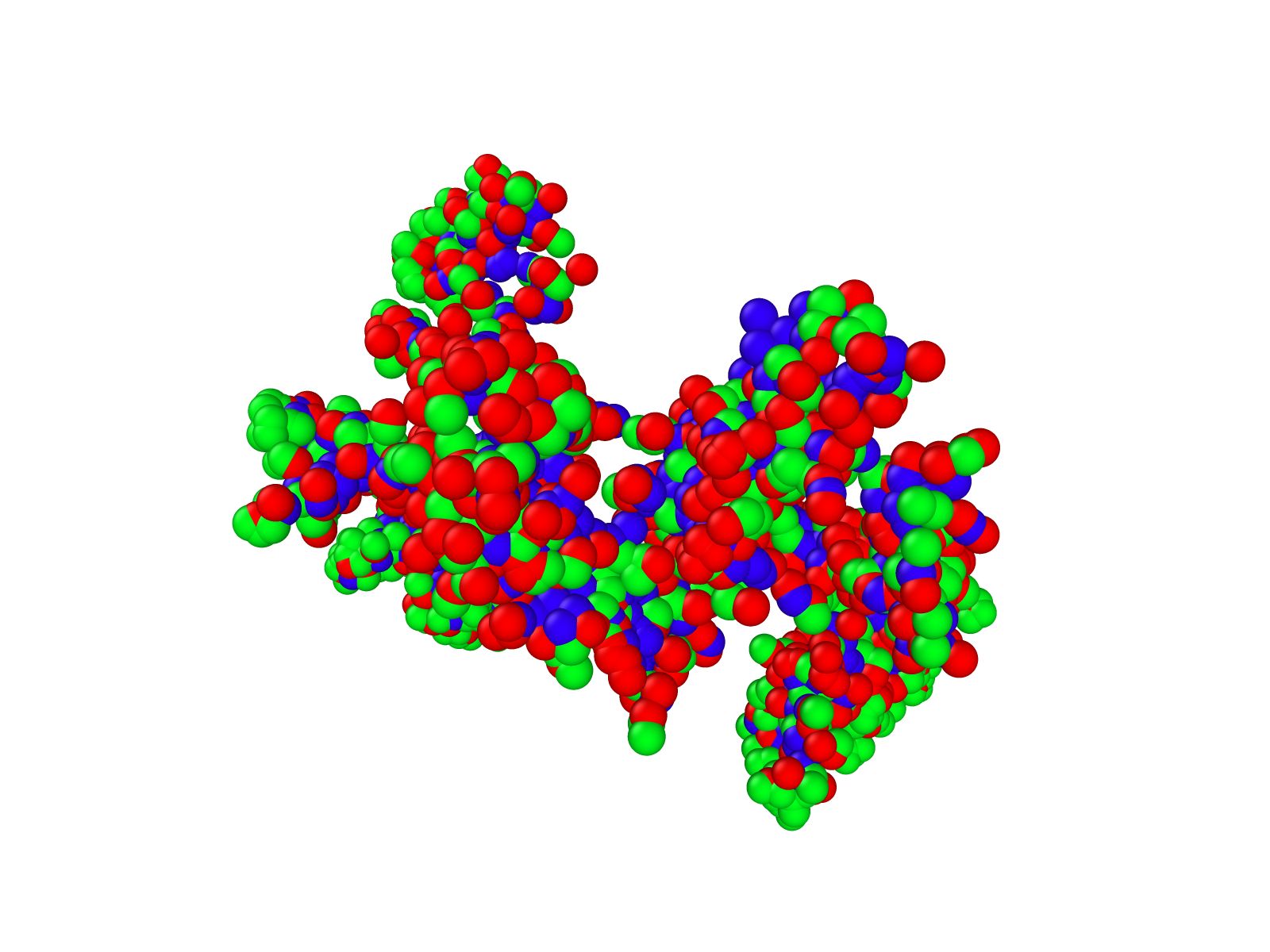}
    \caption{$t=0.3ps$}
  \label{fig:LAMMPS2}
  \end{subfigure}
~
  \begin{subfigure}[b]{0.22\textwidth}
    \includegraphics[width=\textwidth]{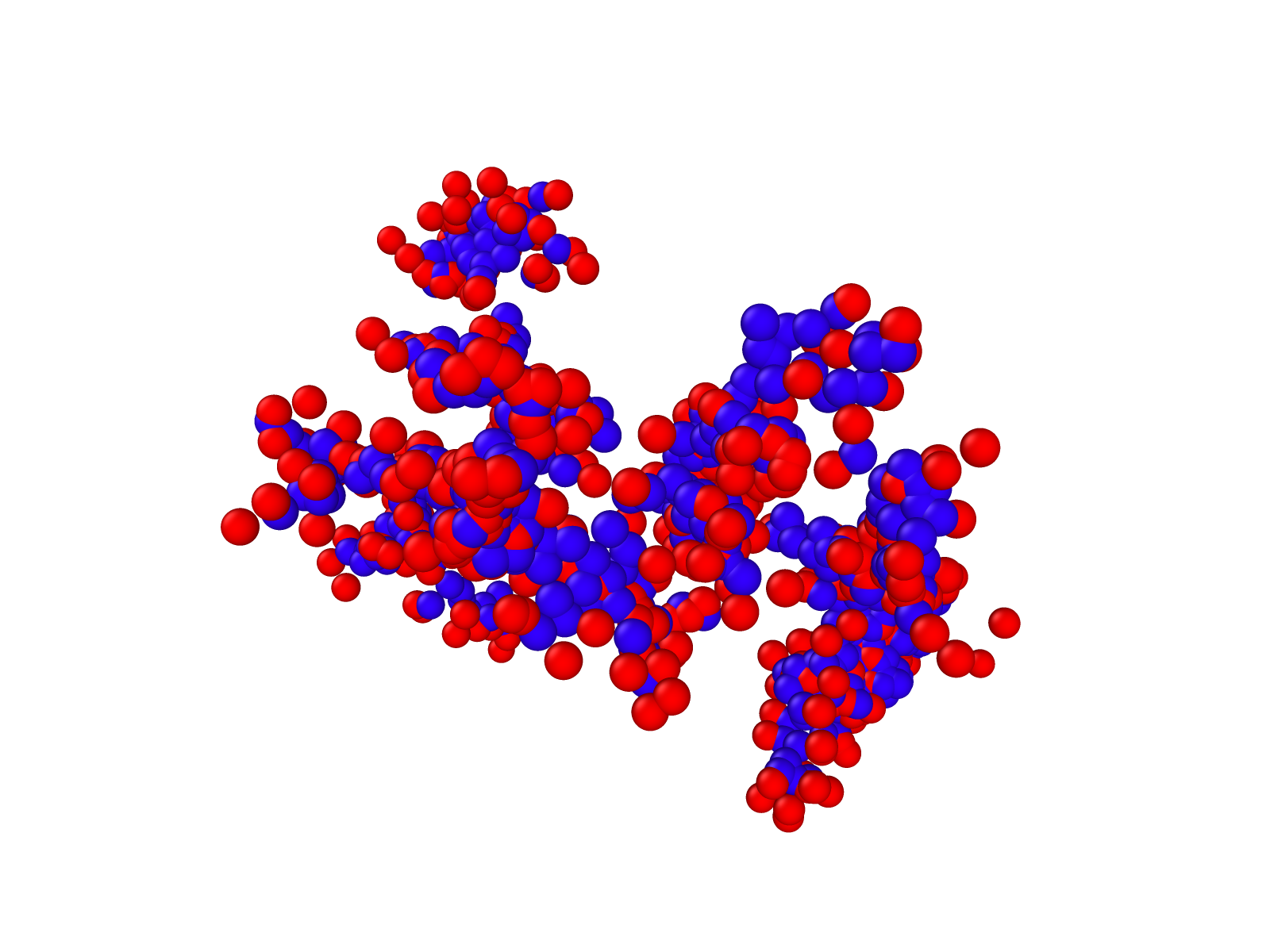}
    \caption{t=11.5ps}
  \label{fig:LAMMPS3}
  \end{subfigure}
~
  \begin{subfigure}[b]{0.95\textwidth}
    \includegraphics[width=\textwidth]{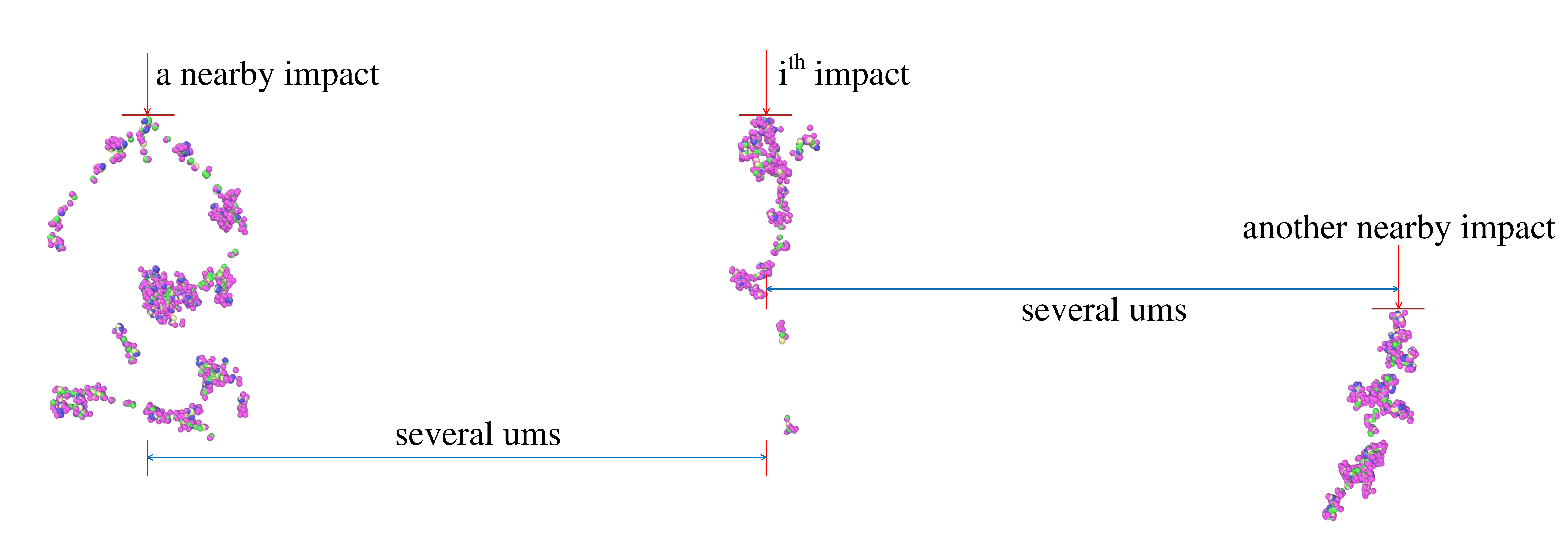}
    \caption{}
  \label{fig:Impact_KMC}
  \end{subfigure}
  \caption{(a)-(d) The fast generation of an initial defect cluster for a PKA of energy $17.1 keV$.
Simulations are made by LAMMPS.
The green balls represent the atoms of high kinetic energy while
the red and blue balls represent static interstitial and vacancy type defects, respectively.
The radiuses of the balls are approximately $5\AA$.
(e) The trajectories of the damage cascades after the impact from the incoming neutrons.
The mean distance between trajectories is also shown.
The figure is derived by combing Monte Carlo, molecular dynamics and kinetic Monte Carlo simulations.}
\label{fig:Impact}
\end{figure*}

The increased $I_{iB}$ of LM324N and LM124 circuits 
are shown as the red and blue curves in Fig.~\ref{fig:Exptresults}. 
The solid symbols represent the mean values of sample splits containing 5 parts and 
the error bars show the standard deviations of the currents.
Explicit increase of the input bias currents are observed on all samples of both types of circuits.
More importantly, the results show apparent divergence among splits exposed at different flux rates.
As shown in the figure,
when the flux decreases from $5\times10^{6}~\text{cm}^{-2}\text{s}^{-1}$ to $5\times10^{5}~\text{cm}^{-2}\text{s}^{-1}$,
for LM324N, the mean value of the excess input bias currents increase monotonously from $0.33~nA$ to $1.35~nA$.
For LM124, the mean value of the excess input bias currents increase monotonously from $1~nA$ to $1.7~nA$.
The explicit divergence of $\Delta I_{iB}$ reveals an ELFS effect of neutron-induced DD.
The increase of the base currents ($I_B$) of GLPNP BJTs irradiated to $1\times10^{10}~\text{cm}^{-2}$ fluence
is plotted in Fig.~\ref{fig:Exptresults} by the green curve.
The results show that,
when the flux decrease from $5\times10^{6}~\text{cm}^{-2}\text{s}^{-1}$ to $5\times10^{5}~\text{cm}^{-2}\text{s}^{-1}$,
the mean value of the excess base currents increase monotonously from $0.5~nA$ to $1.3~nA$.
TCAD simulations confirmed that $\Delta I_B$ shows a linear dependence on the concentration of the defects in silicon bulk
(see Appendix.~\ref{app:TCAD} and Fig.~\ref{fig:GLPNP_Ib}).
Hence, the defect concentraiton follows a similar flux effect as the input bias current or base current.
To investigate the mechanism and characteristics of the effects,
numerical simulations of defects generation have been developed.

\section{Results of the diffusion dynamics limited model}
\label{sec:isolateddefectsmodel}
\subsection{Model details}
\label{chap:discretemodeldetails}

Typical deep-level transient spectroscopy (DLTS) measurement have suggested main (electrically active) defects in silicon induced in DD;
they are
vacancy, $V$, divacancy, $V_2$, vacancy-oxygen complex, $VO$,
vacancy-impurity complex, $VP$ and their charged states~\cite{Irmscher1984,Su1990,Monakhov2002,Fleming2007,Vines2009,Fleming2010}.

\begin{figure}[!t]
\centering
  \begin{subfigure}[b]{0.4\textwidth}
    \includegraphics[width=\textwidth]{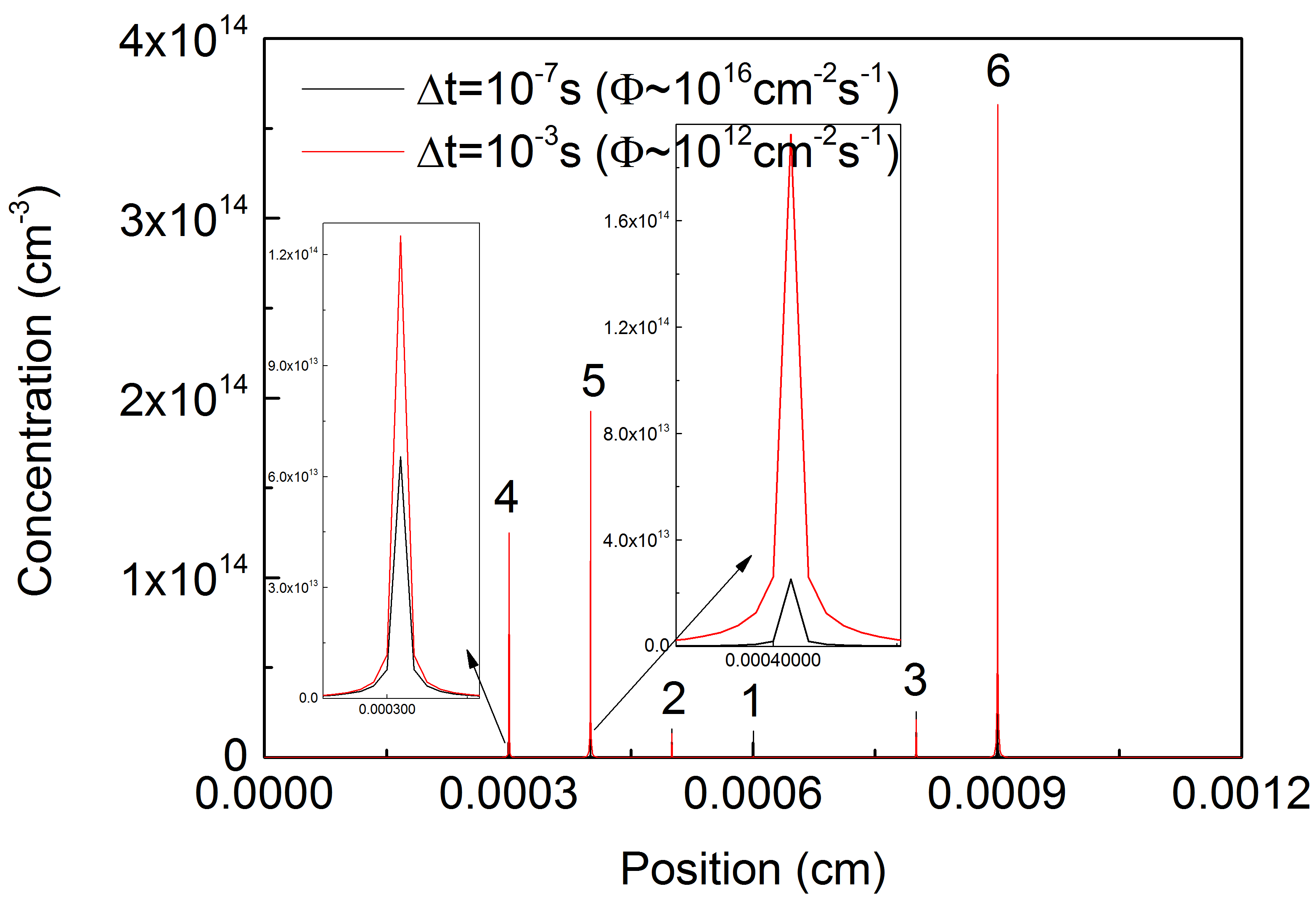}
    \caption{}
  \label{fig:Sim(nodop)_V2}
  \end{subfigure}
~
  \begin{subfigure}[b]{0.4\textwidth}
    \includegraphics[width=\textwidth]{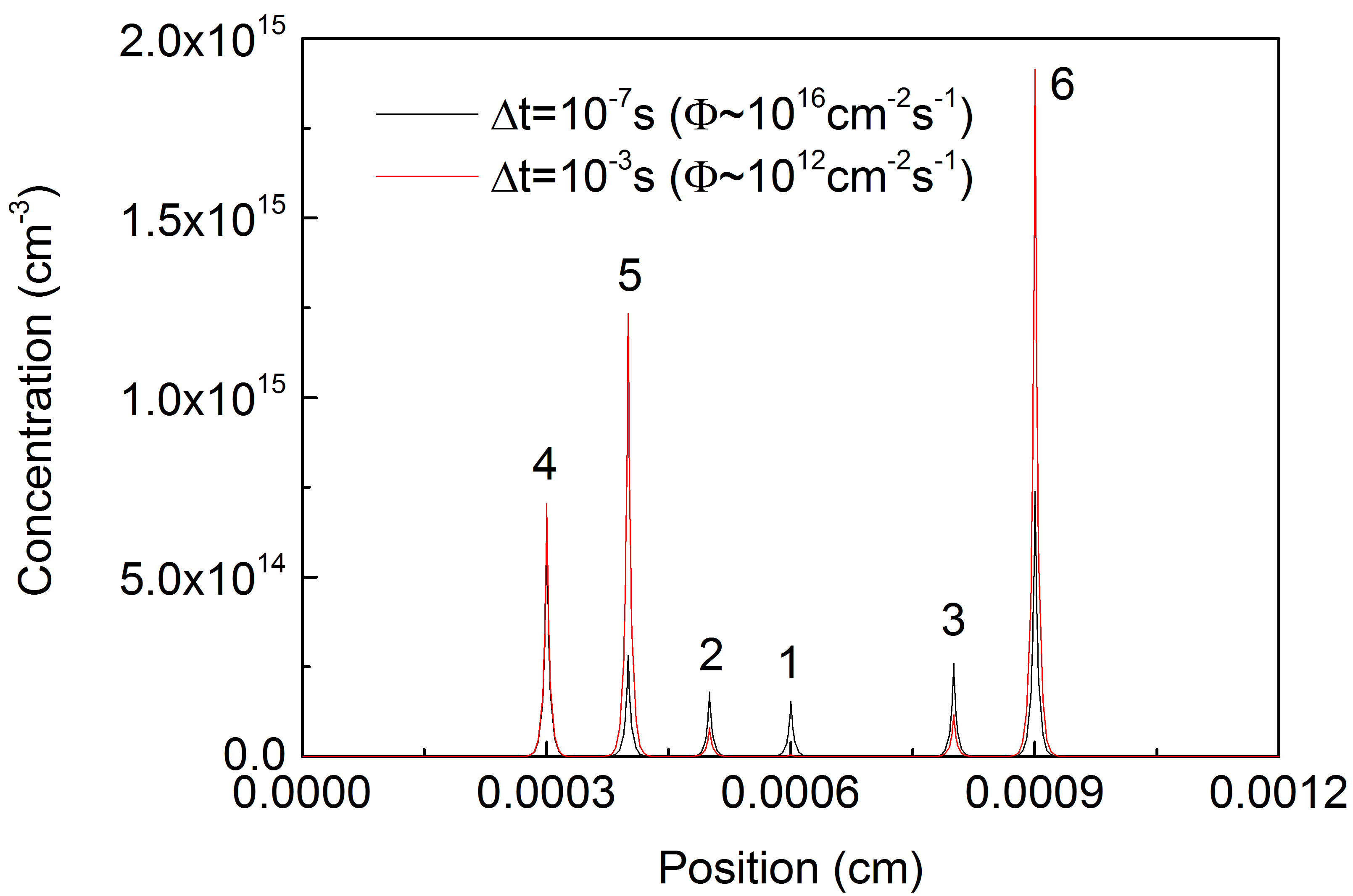}
    \caption{}
  \label{fig:Sim(nodop)_VO}
  \end{subfigure}
~
  \begin{subfigure}[b]{0.43\textwidth}
    \includegraphics[width=\textwidth]{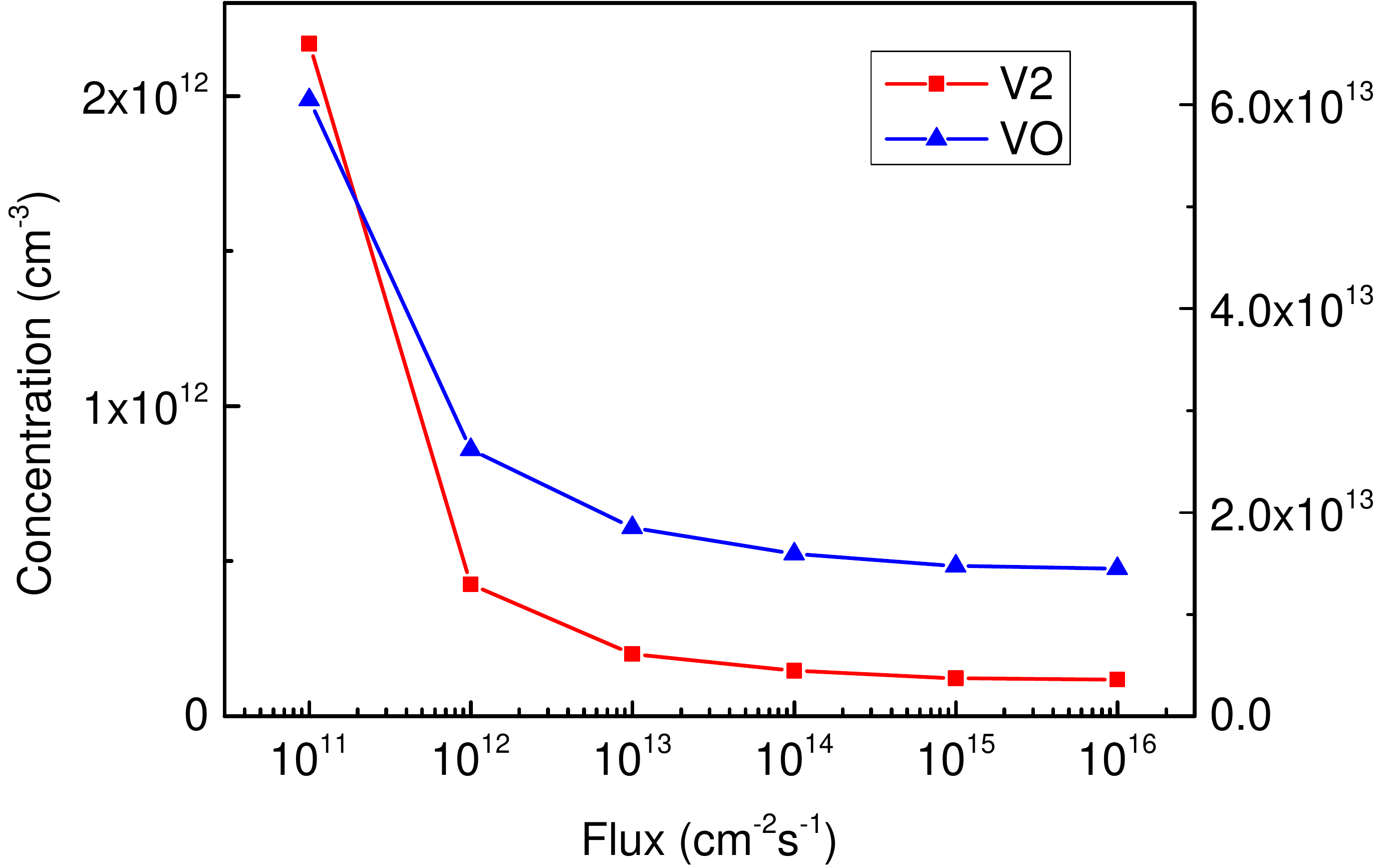}
    \caption{}
  \label{fig:Sim(nodop)_fluxdepend}
  \end{subfigure}
  \caption{(a) Simulated concentration of $V_2$ for two different flux $\Phi$.
For the red dashed curve, $\Delta t=1\times10^{-7}\text{s}~(\Phi\sim10^{16}~\text{cm}^{-2}\text{s}^{-1})$,
for the black solid curve, $\Delta t=1\times10^{-3}\text{s}~(\Phi\sim10^{12}~\text{cm}^{-2}\text{s}^{-1})$.
The orders of the incoming particles is marked near the peaks.
(b) The concentration of $VO$.
(c)Simulated integrated concentration of (left axes)$V_2$ and (right axes)$VO$ for two different flux
as a function of incident neutron flux.}
\label{fig:Sim(nodop)}
\end{figure}

To capture the main characteristics of the damage,
we did one dimensional (1D) simulation of the evolution of the defects.
The simulation method is similar to the one used by Hall\'en \emph{et al}~\cite{Hallen1991}.
Further here, we also take into account the influence of the charge states of the defects
and the charge-promoted hopping of interstitials~\cite{Baraff1984,BarYam1984(1),BarYam1984(2),Car1984,Maroudas1993}.
In the model, the defects of $I$, $V$, $V_2$, $O$, $VO$, 
and their charged states are considered.
The model is rooted on a bimolecular reaction framework and
the reactions among the defects are listed as follows:
\begin{subequations}
\label{eq:whole}
\begin{eqnarray}
I+V \rightarrow \emptyset ~,\label{eq:r0}
\\
I+V_2 \rightarrow V ~,\label{eq:r1}
\\
V+V \rightarrow V_2 ~,\label{eq:r2}
\\
V+O \rightarrow VO ~,\label{eq:r3}
\\
\end{eqnarray}
\end{subequations}
where the subscript $I$ indicates 
an interstitial particle.
The concentration of a particular defect or carrier is determined by
\begin{equation}\label{eq:master}
\frac{\partial[X]}{\partial t} = D_X \frac{\partial^2 [X]}{\partial X^2}-U_X+G_X ,
\end{equation}
where $[X]$ is the density of the ingredient $X$, $D_X$ is the diffusion coefficients,
$U_X$ and $G_X$ are the consumption and generation rate of species $X$, respectively.
The transitions among charged states of the above defects are described through
the capture and excitation of the charge carriers~\cite{Lutz1999},
and the system is thought to be charge neutral as a whole during the simulation.
The model may not cover all possible reactions as we focus primarily on the reactions
having the most prominent influence to the evolution of the displacement defects.
The values of the parameters are chosen from Table 1 in Ref.~\cite{Myers2008}.

The length of the simulated region is defined through a comparison with the experimental configurations.
In our experiments, the total fluence is $1\times10^{10}~\text{cm}^{-2}$,
which results in an average distance of the order of 
$1um$ between collision cascades.
(The generation of a single defect cluster and 
the results of several impacts are illustrated in
Fig.~\ref{fig:Impact}.)
Thus in the 1D approach, the simulated length is set to be $12um$ with total number of impacts equals to 6.
Slight larger margins are left on both edges to relieve the perturbations from the boundary conditions; the vacancy and interstitial concentrations are set to be zero at both edges.
The consequence of each impact particle is treated as newly generated vacancies and interstitials, which distribute
uniformly in a $50\AA$ width region and have a concentration of $1\times10^{18}~\text{cm}^{-3}$.
The impacts are distributed randomly in the simulated region and are separated in time with a constant interval $\Delta t$.

\subsection{Simulation results and its discrepancy from the experimental results}
\label{chap:simulationresults}

Simulated results for silicon is shown in Fig.~\ref{fig:Sim(nodop)}.
In the pristine silicon, the solid defects obtained are mainly $V_2$ and $VO$.
It can be seen from Fig. 4 (a) and (b) that the concentrations of $V_2$ and $VO$ is higher at low flux irradiation (large $\Delta t$)
than the values at high flux irradiation (small $\Delta t$).
The dependence of the integrated concentration of defects on incident flux is plotted in Fig.~\ref{fig:Sim(nodop)_fluxdepend}.
The results 
 show that,
when  a transition flux, $\Phi_t$, of values of $\sim10^{11}\text{-}10^{14}~\text{cm}^{-2}\text{s}^{-1}$,
the integrated number of defects exhibit a clear ELFS effect.
The conversion between $\Phi_t$ and $\Delta t$ is given in Appendix.~\ref{app:fluxcalculation}.

However, in our experiments,
the neutron flux is between $5\times10^5$ and $5\times10^6~\text{cm}^{-2}~\text{s}^{-1}$.
Hence, there is a big discrepancy of transition flux 
between the experiments 
and the simulations. 
Even for the extreme cases, i.e.,
the income particles is not neutron but heavy ions whose energy depositions are more efficient
(assuming that every incoming ion can collide with lattice atoms), 
the value of $\Phi_t$ is no less than $10^{9}\text{-}10^{11}~cm^{-2}s^{-1}$.
This value is still much larger than the fluxes observed in the experiments.
This contrast might be one of the reasons that, 
as claimed in the previous paper~\cite{Hallen1991},
this simulation is only a qualitative support to the experiments and more sophisticated models are required.

\subsection{Reasons for the conflicts}
\label{chap:reasonsofconflicts}
The conflicts between the simulation and experiments stem from the simplifications in the previous models.
In the previous models, the generation processes of defects are treated as
reactions among discrete vacancies, interstitials and impurities in crystalline structures.
However, inside the core of each collision cascade
the initial defects have high concentrations and the amorphous nature of the regions should not be avoided.
In other words, beside discrete defects,
more complicated structures containing several interstitial and/or vacancy-related defects are generated.
As mentioned in Ref.~\cite{Pelaz1999}, after Si implantation,
the interstitials and vacancies survive recombinations are mostly stored in metastable and immobile clusters
($I_{cls}$ and $V_{cls}$).
The metastable structures of interstitial and vacancy-related complexes are also used in
the simulations of Ref.~\cite{Bragado2013, Bragado2008}.
The presence of the intermediate products is also supported by the comparison of characteristic time between
defects diffusing and annealing processes.
Assuming Langevin dynamics is still valid,
the rates of the recombination/annealing process are limited by the mobilities of the defects
and proportional to the concentrations of the reactants.
In our systems, the mobility of the Si interstitial is approximately $10^{-6}\text{-}10^{-4}\text{cm}^2/\text{s}$.
The expected time for a Si interstitial to drift over $50nm$ (which is the approximated size of one damage cascade)
is $10^{-7}\text{-}10^{-5}s$, which is much faster than all observed annealing results after particles implantation
\cite{Sander1966, Srour1970,Srour1972,Srour1973}.
From the above considerations,
within a damage cluster containing high concentration of defects,
the rates of the annealing processes are likely to be dominated by the dissolutions of the defect complexes.
(Also, the interstitials injected from outside the clusters assist the annealing.)
To take into account these considerations and achieve better simulating results,
a modified model containing clustered defects are proposed.

\section{The clusters reordering dynamics limited model}\label{sec:clustermodel}
\subsection{Model details}
\label{chap:clustermodeldetails}

The impact of the incoming particle on Si can create isolated and clustered defects, and
the ratio of the clusters to isolated defects depends on the energies of the primary knock-on atoms (PKAs)~\cite{Srour2003,Srour2013}.
Simulations have shown that a decent amount of defect clusters are generated once the energy of PKA surpasses a threshold of
several keV~\cite{Wood1980}.
Though for simplicity, in many analysis, the clusters are treated as
highly concentrated isolated vacancies, divacancies and interstatials,
e.g. $V$, $V_2$ and $I$~\cite{Myers2008,Hallen1991,Li2014,Fleming2008,Monakhov2002(1)},
the amorphization of the region cause the traditional definitions of vacancies and interstitials making less sense.
Moreover, there is no restrict criteria distinguishing isolated defects and amorphous region,
and in many cases the amorphization of a region of the material is simply defined by
containing defects of concentration overcoming a certain threshold value~\cite{Bragado2013}.
Our particle transport and reaction simulations for 1 MeV incident neutrons performed by GENT4 show that
most PKAs have energies overcoming a few tens or hundreds keV, see Appendix~\ref{app:GENT4}.
The following molecular dynamic calculations identify regions containing high concentrations of clustered defects,
which are constructed within several tens of picoseconds after the impact of a PKA atom,
see Fig.~\ref{fig:Impact}(a)-(d).
The defects of high concentrations can form more complicated defect structures
as described in the last paragraph of Chapter.~\ref{chap:reasonsofconflicts} (e.g. $I_{cls}$, $V_{cls}$, etc.).

The annealing of these defects are different from
the annealing of the isolated interstitials and vacancies of high mobilities.
The reasons are as follows.
1) The properties of the isolated defects derived in crystalline solid are no more valid in the core of the damage cascades;
2) the annealing processes are dominated by the internal reordering mechanisms of the defect complexes.
An evidence is the extraordinary different annealing time of Si samples
bombarded with $1.4~\text{MeV}$ electrons and fission neutrons~\cite{Srour1970,Srour1972,Srour1973}.
In the experiments,
the electron irradiated samples anneal to one-third of their initial damage within $10^{-2}\text{-}10^0~\text{s}$, while
the neutron irradiated samples take much more time, approximately $10^3~\text{s}$, to reach the similar results.
The annealing curves are also different.
The electrons are generally considered to introduce discrete defects distributed throughout the samples, while
the high energy neutrons cause a mass of defect clusters.
Thus the comparisons imply the different annealing processes of the two defect structures.
However importantly, the characteristic annealing time of the 
neutron irradiated damage coincide with
that we observed 
in experiments
($\sim10^3~s$, see Chap.~\ref{chap:simulationresults} and Appendix~\ref{app:GENT4}).
The accordance implies that ELFS effects are results of
the interactions among the annealing of defect clusters,
instead of the simple drift-recombinations of discrete defects.
From these considerations, the problem of the strange characteristic time of simulations in the previous work can be relieved.

\begin{figure}[!t]
\centering
  \begin{subfigure}[b]{0.4\textwidth}
    \includegraphics[width=\columnwidth]{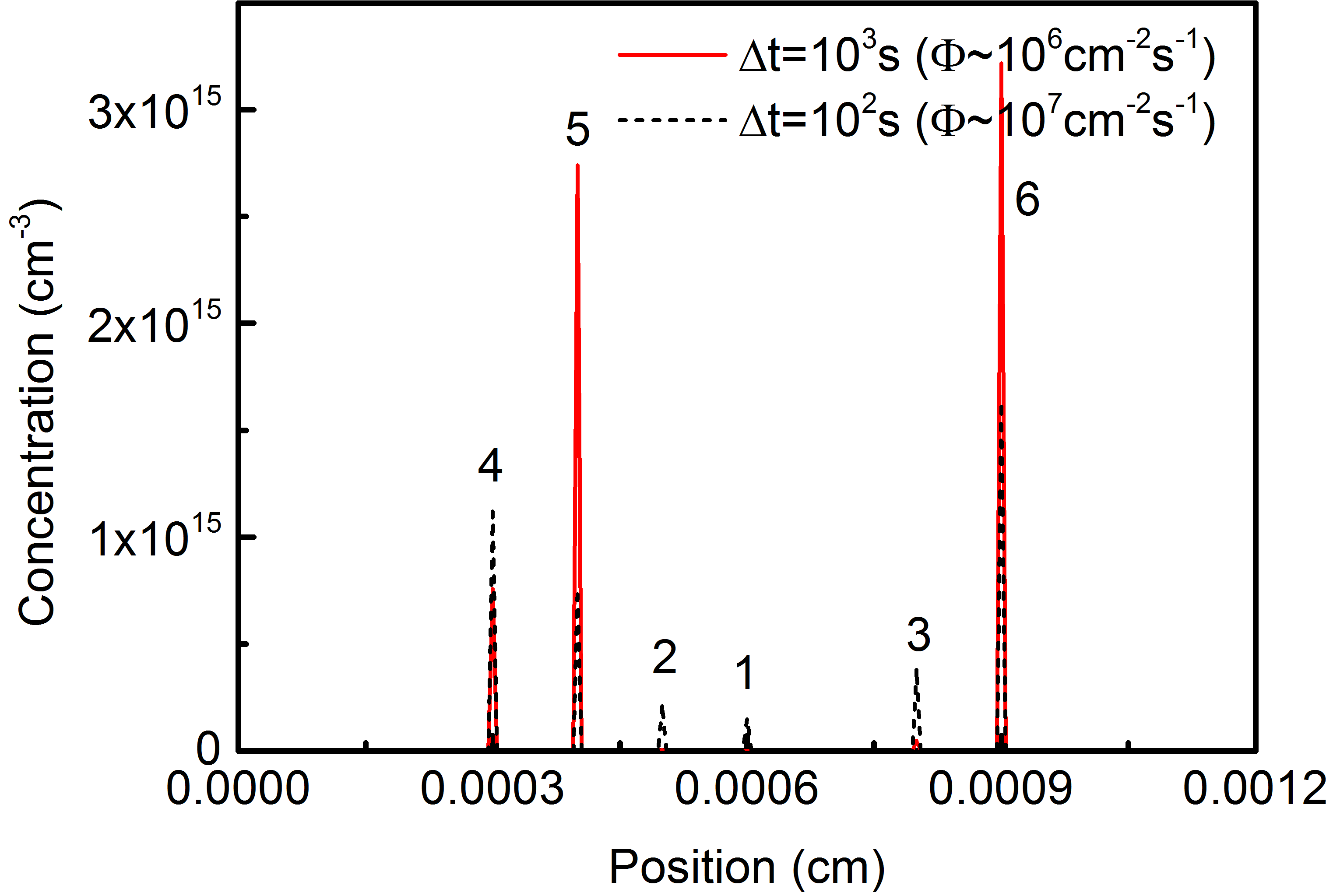}
    \caption{}
  \label{fig:Sim(cluster)_V2}
  \end{subfigure}
~
  \begin{subfigure}[b]{0.4\textwidth}
    \includegraphics[width=\columnwidth]{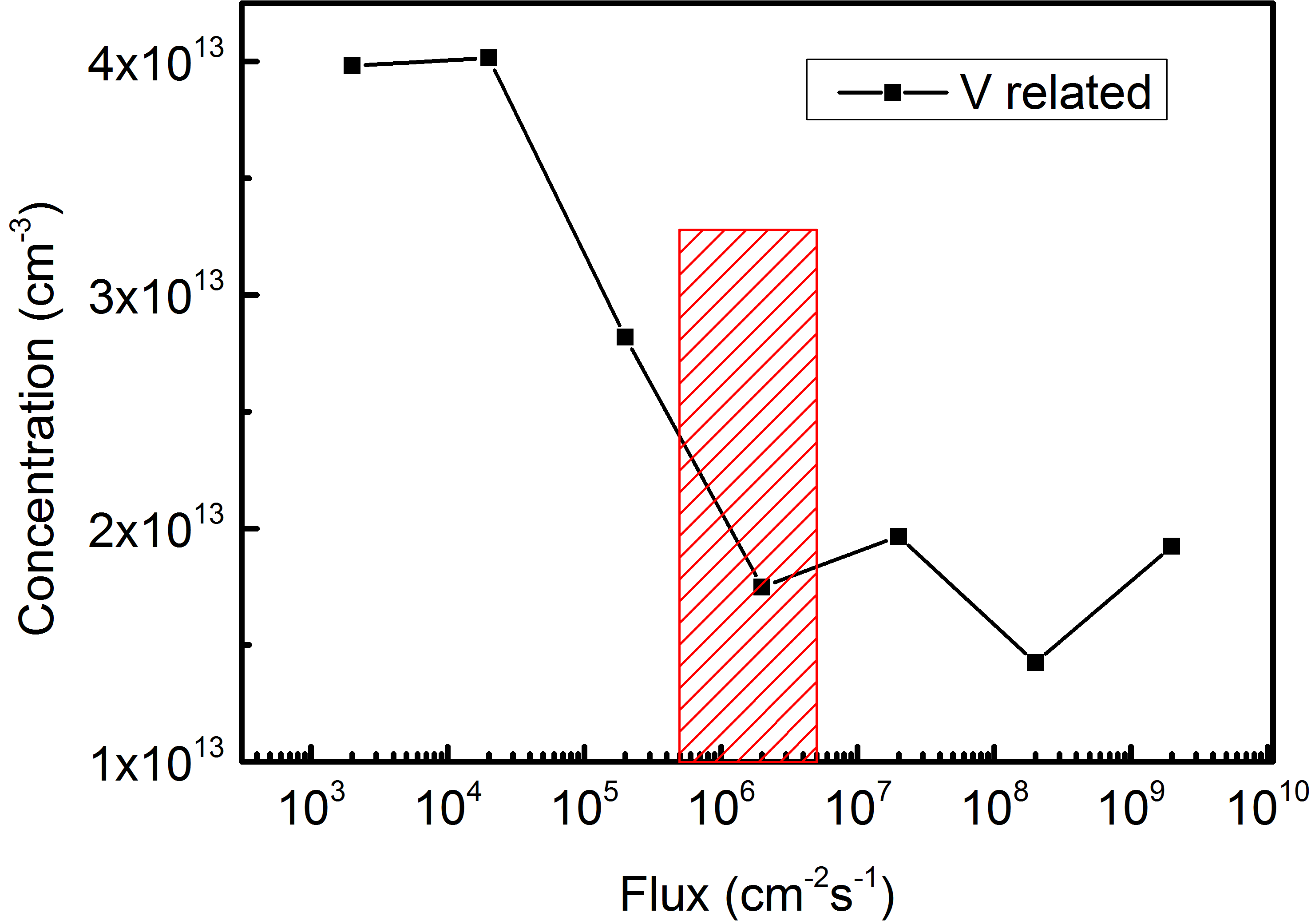}
    \caption{}
  \label{fig:Sim(cluster)_fluxdepend}
  \end{subfigure}
  \caption{(a) Simulated concentration of vacancy($V$) related defects for two different flux using clustered defects model.
For the red dashed curve, $\Delta t=1\times10^{3}s$,
for the black solid curve, $\Delta t=1\times10^{2}s$.
(b) Simulated integrated concentration of defect as a function of incident neutron flux.}
\label{fig:Sim(cluster)}
\end{figure}

Based on the above analyses and the explanations in the last chapter,
a modified model is proposed. 
In this model,
the collision cascade from an incoming neutron is considered to result in complicated defect complexes.
The annealing of the damage is dominated by the reordering of these defect clusters.
The interactions among the clusters are mediated through the particles emitted during the reordering processes.
To explore the primary features of the mechanisms,
numerical simulations of the basic components and reactions have been constructed.
In the calculations,
the defect clusters are defined as a region
containing spatially overlapped immovable interstitial and vacancy-related defects,
$I_{cls}$ and $V_{cls}$.
The results of each collision cascade of incoming neutron are simulated by a defect cluster spanning
$50\AA$ in width containing $1\times10^{18}~\text{cm}^{-3}$ interstitial and vacancy-related defects.
The reordering processes are simulated through the spontaneous decomposition of
the interstitial-related components $I_{cls}$ (emission of Si interstitials) and
the following recombinations of the mobile Si interstitials ($I$) and vacancy-related components ($V_{cls}$).
The decomposition rate is determined by Arrhenius equation $v_dExp(-\frac{E_d}{k_BT})$, and
the energy barrier $E_d$ and the velocity constant $v_d$ are chosen to be $E_d=0.2~eV$ and $v_d=10^{13}/\text{s}$~\cite{Watkins1965}.
The primary reactions in the models are listed below:
\begin{subequations}
\begin{eqnarray}
mI_{cls} \rightarrow (m-1)I_{cls}+I ~,\label{eq:n0}
\\
mV_{cls}+I \rightarrow (m-1)V_{cls} ~,\label{eq:n1}
\\
V+mV_{cls} \rightarrow (m+1)V_{cls} ~,\label{eq:n2}
\end{eqnarray}
\end{subequations}
where $m$ is a counting number.

\subsection{Simulation results}

Results of the simulations are shown in Fig.~\ref{fig:Sim(cluster)}.
The region of calculation is $12~\mu m$ width and $6$ successive impacts are used.
The time interval $\Delta t$ varies from $10^1~s$ to $10^4~s$.
The configurations designed in the 1D simulation is to best mimic the real 3D experimental configurations,
as described in Chap.~\ref{chap:discretemodeldetails}.

\begin{figure}[!t]
\centering
  \begin{subfigure}[!t]{0.36\textwidth}
    \includegraphics[width=\columnwidth]{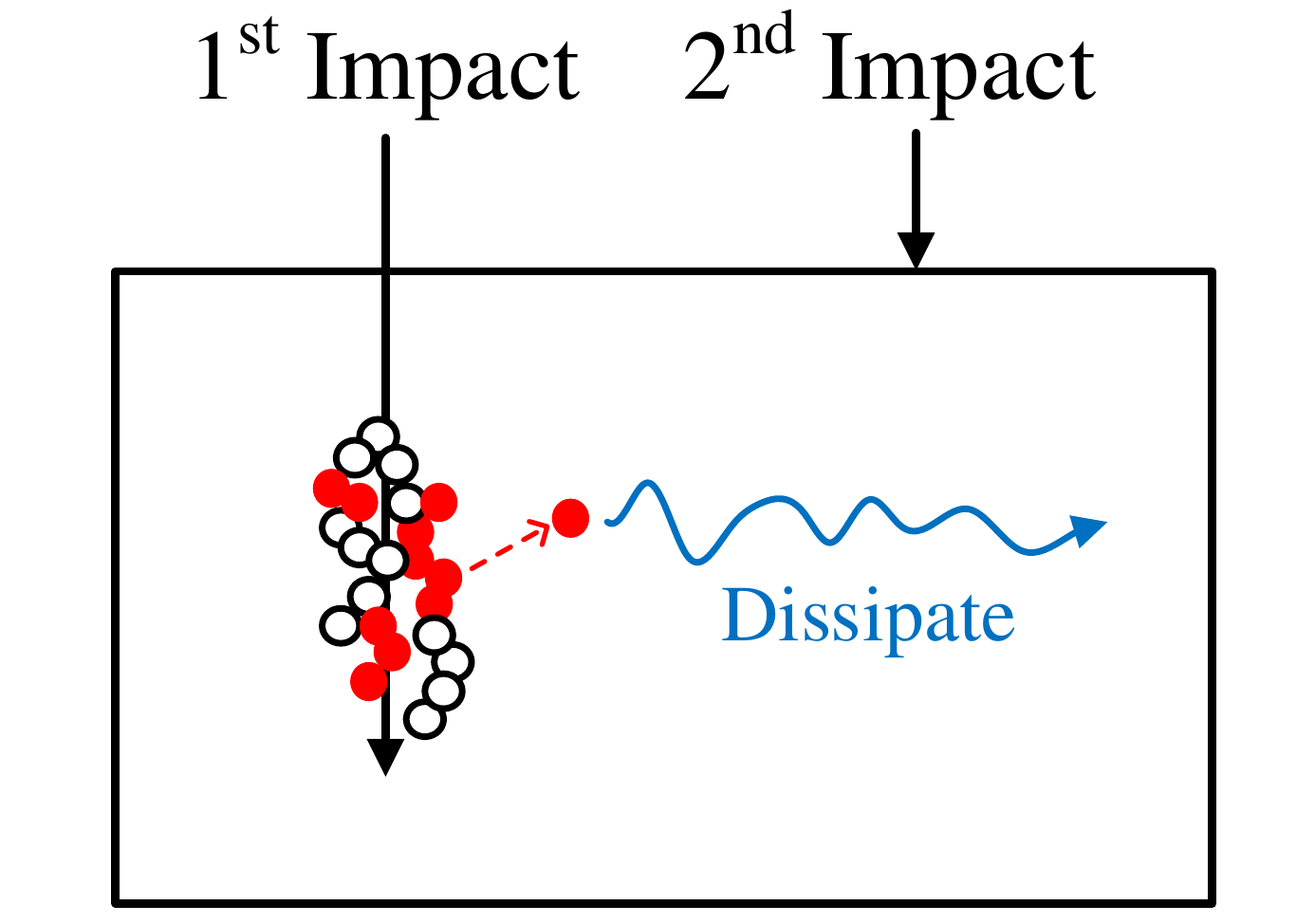}
    \caption{}
  \label{fig:mechanism_interactions_slow}
  \end{subfigure}
~
  \begin{subfigure}[!t]{0.36\textwidth}
    \includegraphics[width=\columnwidth]{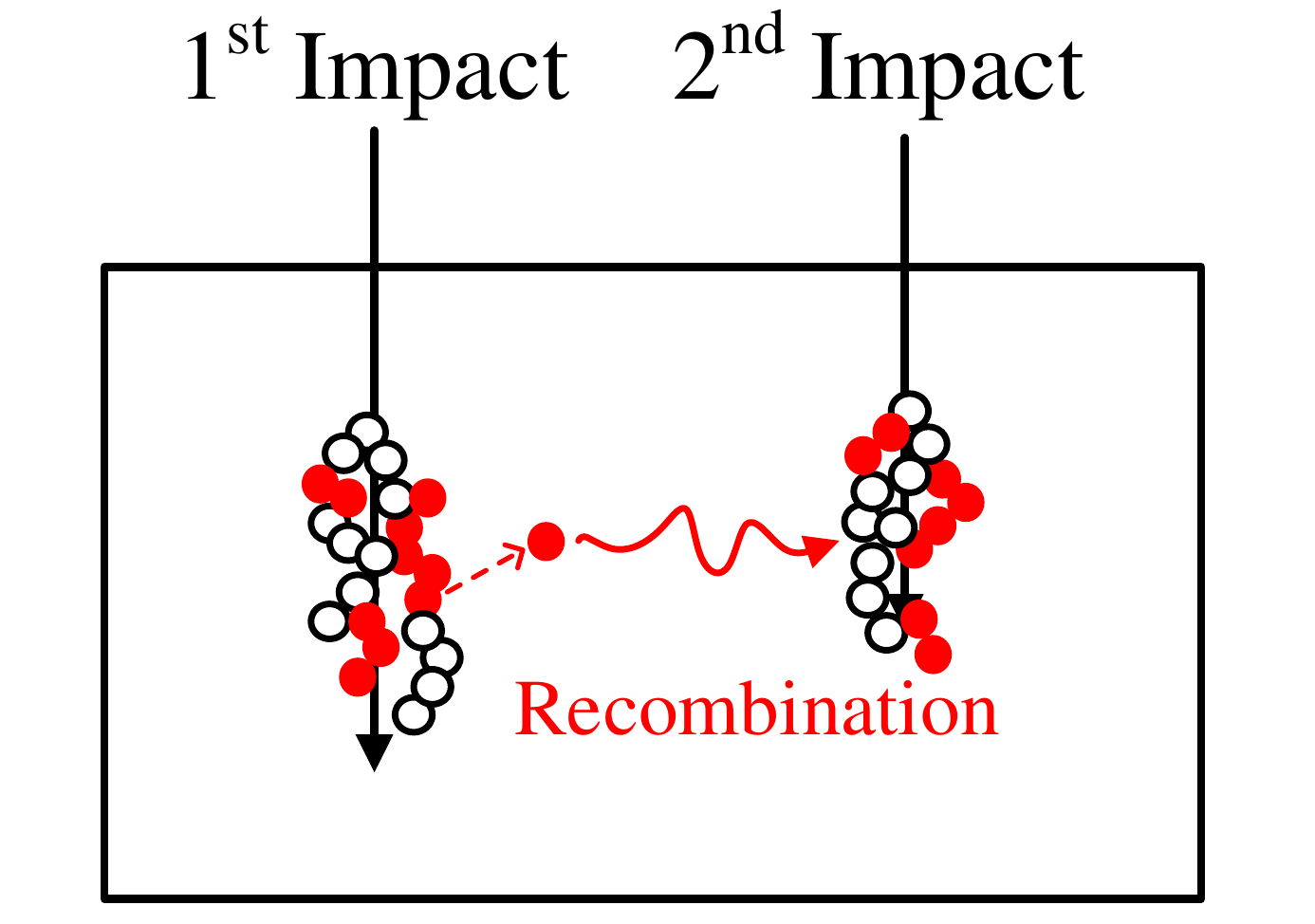}
    \caption{}
  \label{fig:mechanism_interactions_fast}
  \end{subfigure}
  \caption{Schematic show of the difference between the annealing processes of
(a) high flux (small $\Delta t$) irradiation and (b) low flux (large $\Delta t$) irradiation.}
\label{fig:mechanism_interactions}
\end{figure}

In Fig.~\ref{fig:Sim(cluster)_V2},
the concentrations of vacancy related defects are higher at lower flux irradiation (big $\Delta t$) than
at higher flux irradiation (small $\Delta t$).
The integrated numbers of defects are plotted in Fig.~\ref{fig:Sim(cluster)_fluxdepend} as a function of the flux.
An inverse S-shaped curve is derived, which contains two near-flat plateaus at low and high flux edges and 
a fast changing region in the middle.
The results are similar to those given by the discrete defects model
(Fig.~\ref{fig:Sim(nodop)_fluxdepend} in Chap.~\ref{sec:isolateddefectsmodel}), but
have remarkably lower sensitive flux region (or larger $\Delta t$).
The transition flux 
show better agreements with the experimental results.
It explains the experiences that why we only observe explicit ELFS effects for neutrons bombardment
at low flux ($<10^7~\text{cm}^{-2}\text{s}^{-1}$).
The model also predicts a saturation of DD for very low flux irradiations.
Though many merits the model possesses,
more sophisticated models 
are still required and the concerns will be discussed in Chap.~\ref{sec:discussion}.

\subsection{Mechanisms for the neutron-induced ELFS effect}
\label{chap:mechanisms}

The way how the flux rate effects stem from the interactions between defect clusters is illustrated in Fig.~\ref{fig:mechanism_interactions}.
For low flux irradiations, the time intervals $\Delta t$ between the sequent impacts of incoming neutrons are large.
Before the following impacts happen,
the defect cluster in an impact has enough time to anneal.
During the processes, the metastable structures decompose and emit mobile particles (mainly Si interstitials),
which can traverse towards the edges and
be absorbed by the surface or re-merge/disappear somewhere in the lattice \cite{Cowern1999(2),Jung2004,Jung2004,Agarwal1997,Vuong2000}
(see Fig.~\ref{fig:mechanism_interactions_slow}).
Thus the damage cascades of subsequent impacts do not feel the influence of the previous impacts.
While for high flux irradiations, the time intervals between the sequent impacts are small,
and 
are not enough for the annealing of each isolated defect cluster.
In this limit, the particles ejected from the previously created defect clusters gain increased possibilities
to encounter the subsequent defect clusters before disappearing (see Fig.~\ref{fig:mechanism_interactions_fast}).
The effect enhances the efficiency
of the reordering processes of the defect clusters,
which results in a reduced number of defects.
This is the experimentally observed ELFS effect.

\begin{figure}
  \includegraphics[width=0.45\textwidth]{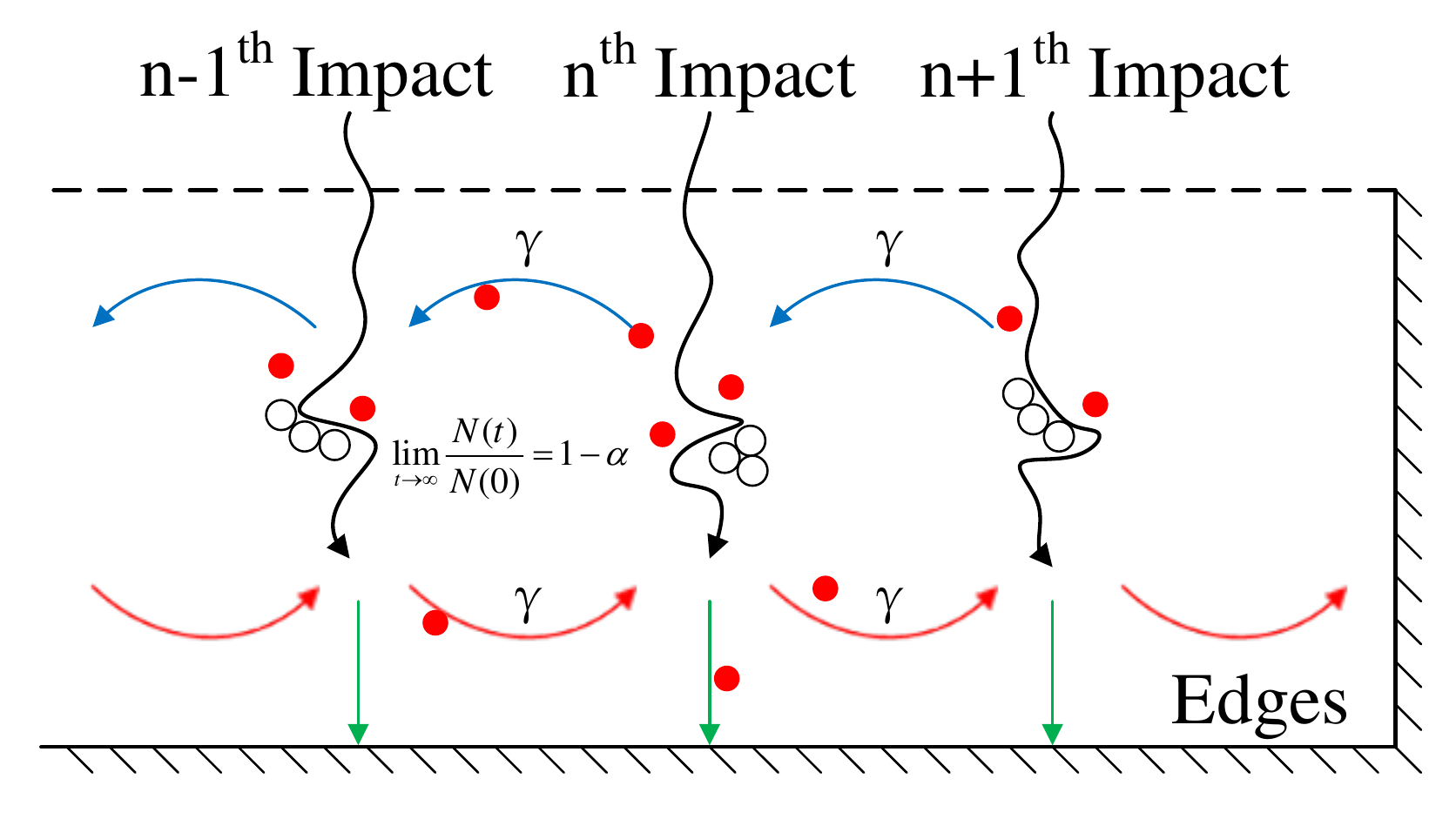}
  \caption{Simplified analytic model describing the interactions between defect clusters of successive neutron impacts.}
  \label{fig:model_analytic}
\end{figure}

\begin{table}
\centering
\caption{Parameter list}
\begin{tabular}{|c|>{\makecell*[c]}c|c|c|}
\hline
Parameters & $\alpha$ & $\tau$ & $n^s$ \\
\hline
LM324 & 0.954 & $1.75\times10^3s$ & 5 \\
\hline
LM124 & 0.82 & $1.5\times10^3s$ & 5 \\
\hline
GLPNP & 0.9 & $3\times10^3s$ & 6.1 \\
\hline
\end{tabular}\label{tab:model_parameters}
\end{table}

\section{Parametric model}\label{sec:Parametricmodel}

The mechanism is further extracted to construct a simplified model illustrated in Fig.~\ref{fig:model_analytic}.
The primary parameters of the mechanisms 
are extracted and tabulated
in Table.~\ref{tab:model_parameters}.
The three main ansatz of the model are:

1.~Each isolated defect cluster can anneal to a ratio, $1-\alpha$ of its initial size.
The reordering process has an average characteristic time $\tau$.

2.~The primary mobile elements are Si interstitials, which lose a part of their amount each time
when they encounter another defect cluster.
As the concentration of defects increases, the capturing processes tend to saturate.
For each given system,
a parameter $n^s$ can be specified to characterize the change of efficiency of the capturing processes.
The reduced interstitials through the capture processes contribute to the increased number of the annealed defects.

3.~The mobile particles, once reach the edges, will be absorbed and no more be rejected back.

The flux rate, which determines the average intervals $\Delta t$ between successive collision cascades,
determines the number of the defect clusters that can be encountered by the Si interstitials
emitted from the previously created defect clusters before disappearing.

The model gives an analytical solution of the dependence of the enhancement factors (EF) on irradiation flux.
The remnant number of defects $n^{rem}$
(in measure of a defect cluster of standard size, e.g. $n^{rem}=0.5$ if a cluster annealed to half of the size of a standard defect cluster),
after a given fluence of irradiation $\Phi$, is given by the following equation
(Eq.\ref{eq:parametricmodel}).
The detailed derivative is given in Appendix~\ref{app:parametricmodel}.

\begin{widetext}
\begin{equation}\label{eq:parametricmodel}
n^{rem}=n^{s}\text{ln}\left(\frac{1}{-1+\alpha}\left(e^{-n^{r,eff}/n^{s}}\alpha+e^{-n^{tot}(-1+\alpha)/n^{s}}\left(-1+\alpha-e^{-n^{r,eff}/n^{s}}\alpha\right)\right)\right) ~,
\end{equation}
\end{widetext}
In the equations, the meanings of $n^s$ and $\alpha$ have been indicated above.
$n^{r,eff}$ calculates the effective number of collision cascades involved in
capturing escaped particles from one damage cascade.
\begin{equation}\label{eq:parametricmodel_1}
n^{r,eff}=\gamma\kappa\tau e^{-1/(\gamma\kappa\tau)} ~,
\end{equation}
where $\gamma$ is the flux of the irradiation and $\tau$ is the characteristic time of self annealing of a single damage cascade.
The total amount of collisions is measured by $n^{tot}$,
\begin{equation}\label{eq:parametricmodel_2}
n^{tot}=\kappa \Phi ~,
\end{equation}
where $\kappa$ measures the probability of the collisions taking place for a unit fluence of incoming particles,
which has the unit of $cm^2$.
The DD is given by 
\begin{equation}\label{eq:parametricmodel_3}
\Delta I_B=k n^{rem} ~,
\end{equation}
where $k$ is a constant connecting the number of defects with the parametric change of the devices.

\begin{figure}[!t]
\centering
  \begin{subfigure}[b]{0.4\textwidth}
    \includegraphics[width=\textwidth]{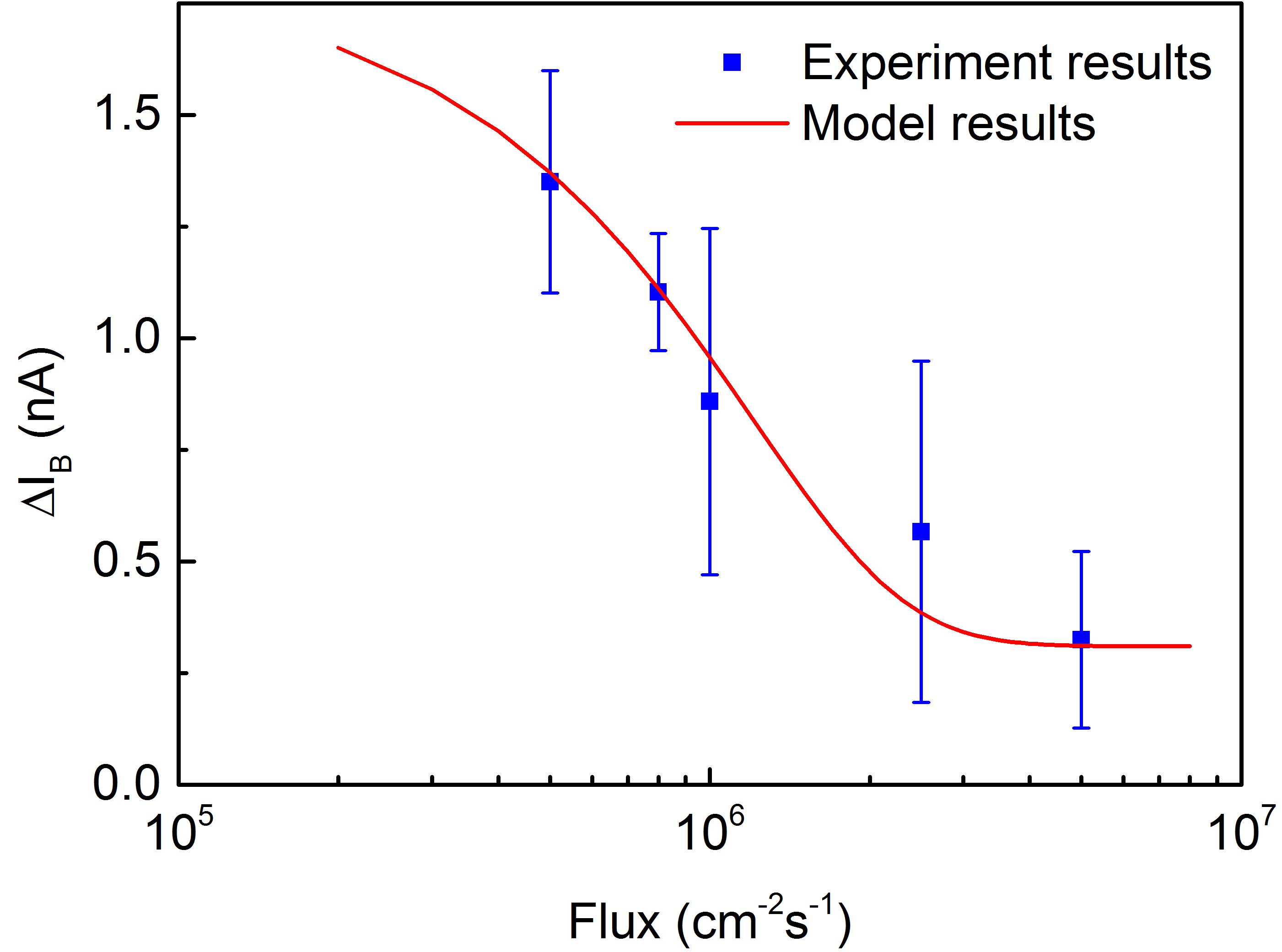}
    \caption{}
  \label{fig:FittingLM324N}
  \end{subfigure}
~
  \begin{subfigure}[b]{0.4\textwidth}
    \includegraphics[width=\textwidth]{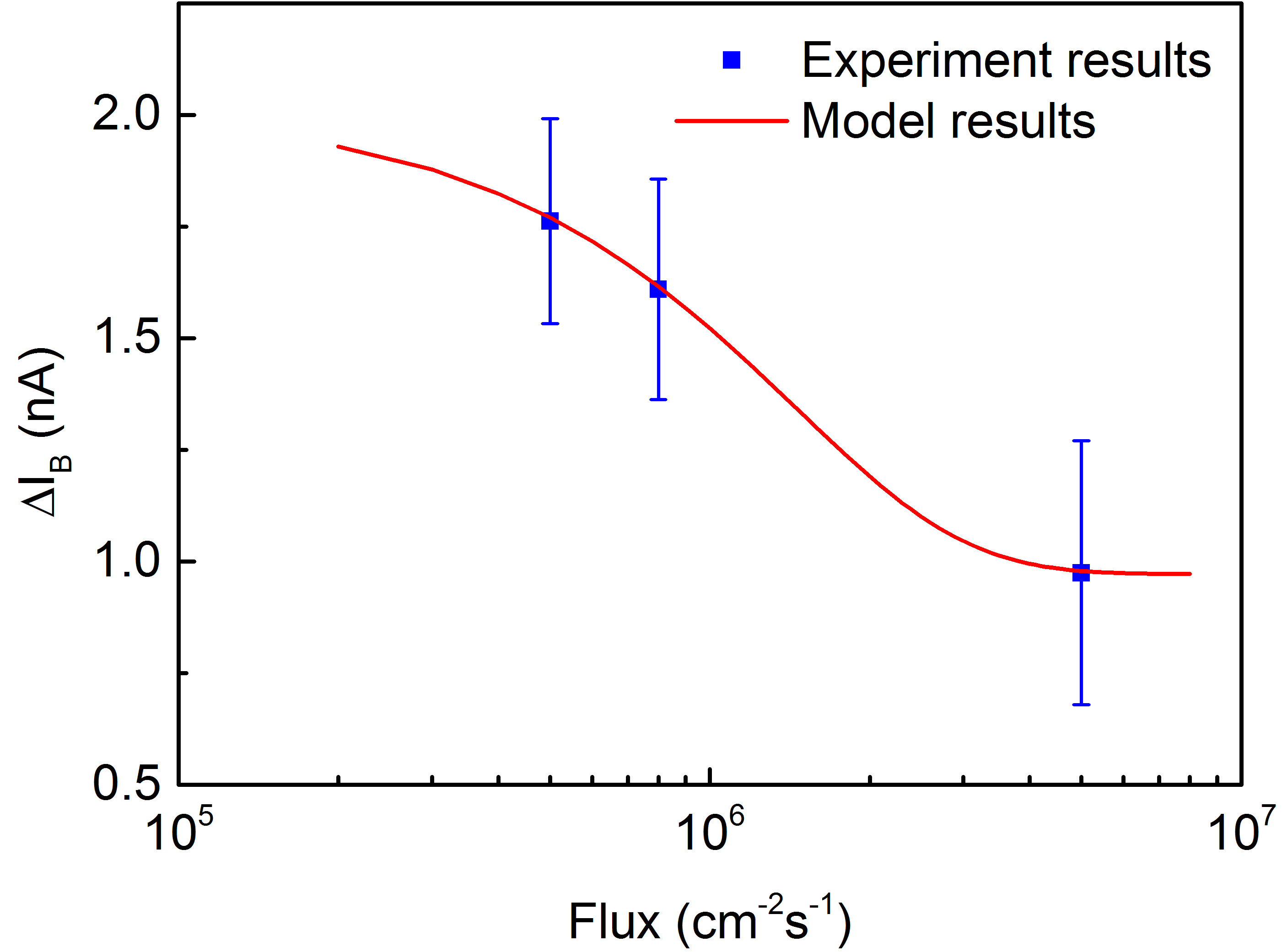}
    \caption{}
  \label{fig:FittingLM124}
  \end{subfigure}
~
  \begin{subfigure}[b]{0.4\textwidth}
    \includegraphics[width=\textwidth]{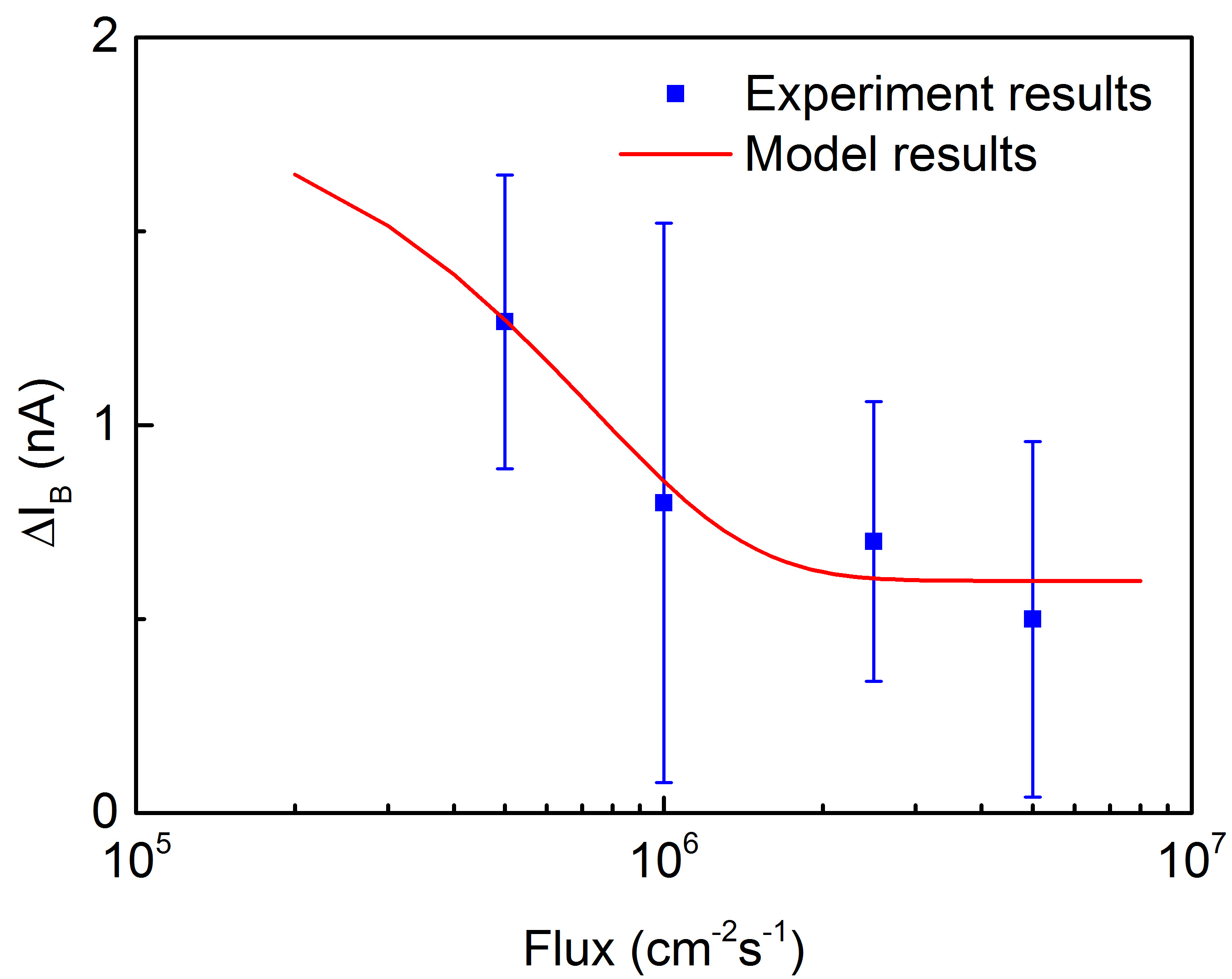}
    \caption{}
  \label{fig:FittingGLPNP}
  \end{subfigure}
  \caption{After $10^{10}cm^{-2}$ fluence of neutron irradiation for (a)LM324N, (b)LM124, (c)GLPNP.
Solid lines are the calculated $\Delta I_B$ derived from Eq.~\ref{eq:parametricmodel} and
points are the experimental results.
The parameters chosen in the calculations are:
(LM324N)$\tau=1.75\times10^3s$, $\kappa=5\times10^{-9}cm^2$, $\alpha=0.954$, $n^{s}=5$, $k=0.135nA$.
(LM124)$\tau=1.5\times10^3s$, $\kappa=4\times10^{-9}cm^2$, $\alpha=0.82$, $n^{s}=5$, $k=0.135nA$.
(GLPNP)$\tau=3\times10^3s$, $\kappa=5.3\times10^{-9}cm^2$, $\alpha=0.9$, $n^{s}=6.1$, $k=0.113nA$.}
\end{figure}

Following the discussions in the previous chapters
(Chap.~\ref{chap:simulationresults}, \ref{chap:clustermodeldetails} \& Appendix~\ref{app:fluxcalculation}),
the experimental results suggest the annealing time of neutrons bombardment 
is of magnitude of $10^3s$ and
the probability of the collisions 
is approximately $1-10$ times per $10^9\text{cm}^{-2}$ fluence.
Thus the reasonable values of the parameters would be $\tau\sim10^3s$ and $\kappa^{-1}\sim10^8-10^9\text{cm}^{-2}$.
The comparison between calculated values and experimental results of the damage of LM324N
is plotted in Fig.~\ref{fig:FittingLM324N}.
The parameters used are given in the figure caption.
The values of $\Delta I_B$ calculated through the parametric model (Eq.~\ref{eq:parametricmodel}-\ref{eq:parametricmodel_3}) show
good agreements with the experimental results.
In the simulation, we noticed that the damage (represented by $\Delta I_B$) has an evident flux dependence
in the range of $5\times10^5-5\times10^6\text{cm}^{-2}\text{s}^{-1}$.
When the flux is much higher than $5\times10^6\text{cm}^{-2}\text{s}^{-1}$ or much lower than $5\times10^5\text{cm}^{-2}\text{s}^{-1}$,
the flux sensitivity becomes trivial.

The reasons can be explained as follows.
At very low flux,
the post-directional interactions 
among defect clusters
become very weak.
In this limit, further decreasing of the flux rates will not reduce the recombinations effectively.
At very high flux,
on the contrary, the post-directional interactions are very efficient to the extent of nearly saturation.
In this limit, further increasing of the flux rates will not enhance the recombinations effectively.
Therefore, in both extremes,
the damage show little sensitivities to the flux rates.

Similar calculations have also been made for LM124 and GLPNP.
The results are shown in Fig.~\ref{fig:FittingLM124} and \ref{fig:FittingGLPNP}, respectively.
Both results also show good agreements between the modeling results and the experimental results.

\begin{figure}[!b]
  \includegraphics[width=0.4\textwidth]{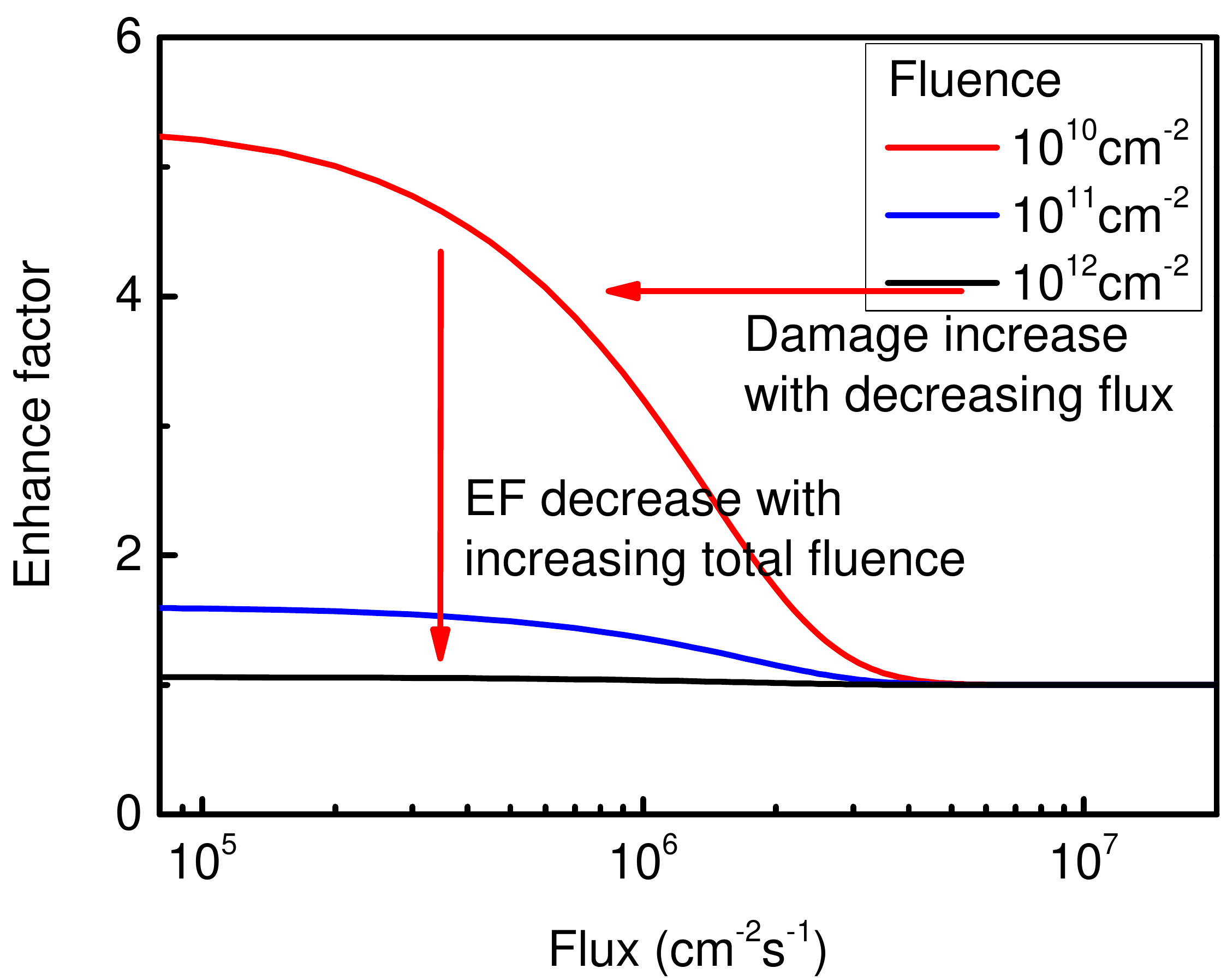}
  \caption{The dependence of the enhancement factor on flux at total fluence of (red)$10^{10}cm^{-2}$, (blue)$10^{11}cm^{-2}$ and (black)$10^{12}cm^{-2}$.}
  \label{fig:fluencedependence}
\end{figure}

\begin{figure}[!t]
  \includegraphics[width=0.4\textwidth]{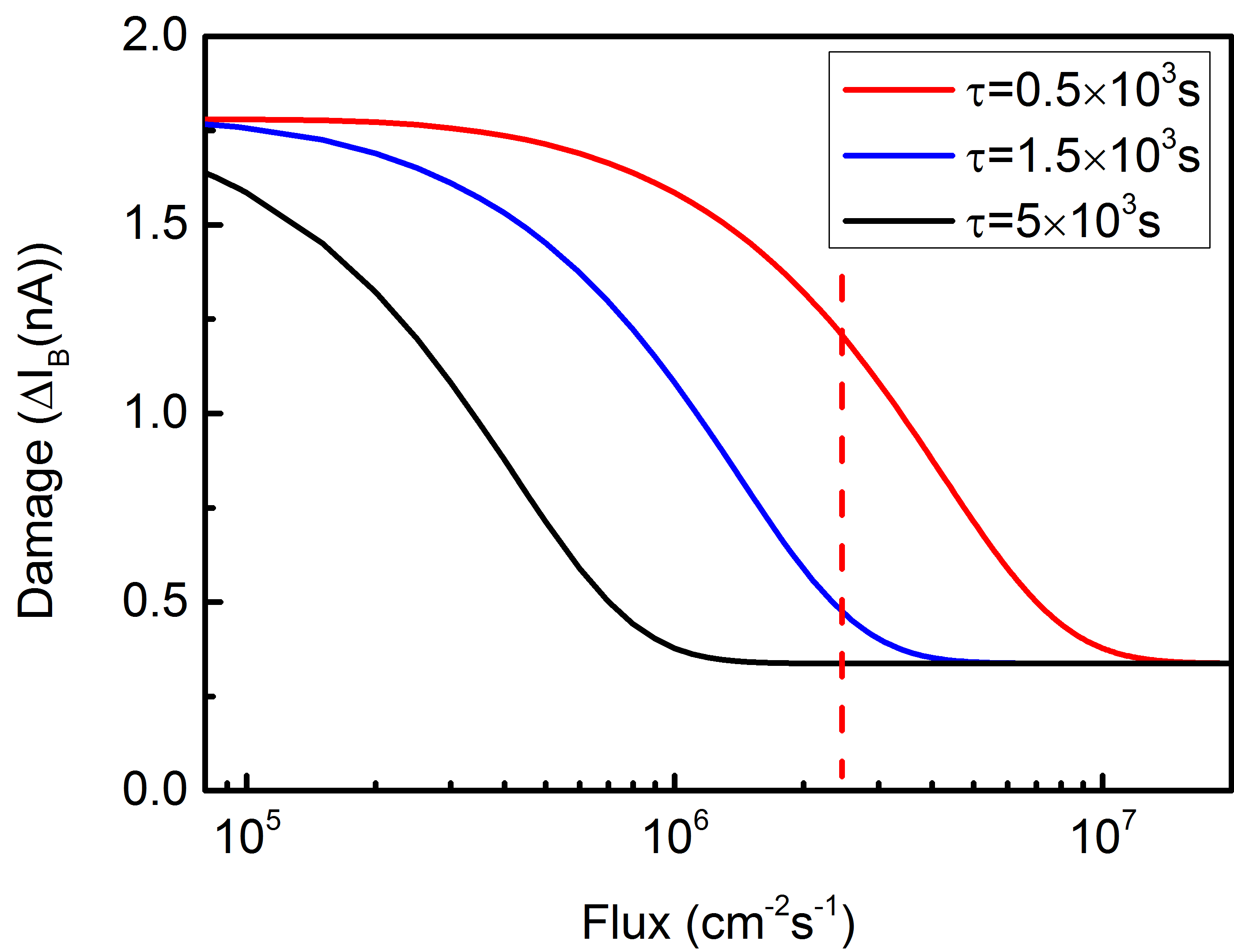}
  \caption{The dependence of damage on flux at fixed total fluence of $10^{10}cm^{-2}$ for different temperatures:
(red) high temperature with $\tau=0.5\times10^3s$,
(blue) middle temperature with $\tau=1.5\times10^3s$,
(black) low temperature with $\tau=5\times10^3s$.
For fixed flux (e.g. $2.5\times10^6~cm^{-2}s^{-1}$, indicated by red vertical line),
the damage increases with increasing temperature (decreasing $\tau$).
The parameters used in the figures are $\alpha=0.95$, $\kappa=5$, $n_s=5$, $k=0.135nA$.}
  \label{fig:temperaturedependence}
\end{figure}

\subsection{Prediction of fluence dependence of the flux effect}

In the above section,
we have demonstrated the matches between the results of the parametric model and the experimental results.
Moreover, besides the flux rate dependence of the EF, 
the model also predicts a nontrivial dependence on the fluence. 
The results are
illustrated in Fig.~\ref{fig:fluencedependence}.
The results show that, with the increase of the total fluence, EF decreases.
When the fluence increases to $10^{12}\text{cm}^{-2}$, the flux effect almost disappears.
This characteristic explains the phenomenon that
the flux effect of neutrons bombardment was not observed in previous experiments of high fluence ($>10^{12}\text{cm}^{-2}$).
The reasons come from the accumulations of the defects in the system.
After each damage cascade, the new generated Si interstitials can annihilate with defects
created before or after themselves' generation.
For low fluence cases, the post-directional annealing is important comparing with pre-directional annealing.
When the fluence increase, more defects are accumulated which leads to the increase of the weight of the pre-directional annealing.
However, the pre-directional annealing does not have flux dependence because
the strength of the mechanism is dominated by the number of the existing defects rather than
the accurate time they were created.
Thus in high fluence irradiation experiments, the flux dependent effects are suppressed.

\subsection{Prediction of temperature dependence of the flux effect}

The temperature is generally recognized to influence the annealing rate of defects.
In the parametric model, the influences of the temperatures are treated as the changes of the characteristic time $\tau$ which
measures the annealing rates of defects\cite{Srour1970,Vook2006}.
Higher temperatures accelerate the annealing processes hence result in smaller $\tau$'s.
The calculated results of damage at different temperatures are plotted in Fig.~\ref{fig:temperaturedependence}.
From the figure we can see that,
with the increase of the temperatures (decrease of $\tau$),
the damage curves shift to the right.
If we choose a fixed flux as indicated by the red vertical line for an instance in the figure,
the damage is found to increase with increasing temperature.

\section{Discussion}
\label{sec:discussion}

In the continuous model simulations, though the results show qualitative agreements with the experiment results,
there are some discrepancies of the amplitudes between the simulated results and the experimental observations.
Two concerns may influence the accuracy of the results.

1.~The inconsistence between the continuous modeling used in simulations and
the discrete nature of the real dynamics of the defects.
The validity of the continuous approach of the molecular dynamics of a discrete system requires that
the mean free path (MFP) of the molecular is much smaller than the characteristic scale of the system.
In our systems, a defect cluster has a very narrow radius ($\sim10nm$).
The length may be comparable with the MFP of mobile defects.
In the space among the clusters,
the concentrations of the defects are low and the MFP of mobile defects may be comparable with the size of the whole system.
Therefore,
the continuous modeling may not be the best simulation method in mimicking the dynamics of the defects evolution
and its accuracy is degraded.

2.~In the numerical simulations, the approaches of the reordering mechanisms within defect clusters are too simple.
When describing the interactions between interstitial and vacancy-related defects inside and outside the defect clusters,
the modeling uses the same diffusion coefficient of mobile particles and rate of reactions.
The approaches require the validity of Langevin dynamics inside defect clusters which may not be exact in practice.
If non-Lagevin dynamics dominate inside defect clusters,
the defect clusters would behave like black bodies which absorb incoming particles with little reflections.
Particles leave the clusters through emission mechanisms.
New parameters are then required to characterize the recombination processes.
An analogy is given in Ref.~\cite{Bubon2016} when describing the columnar recombination of x-ray generated electron-hole pairs.
To fit the experimental results, the author coupled the contributions of Langevin and non-Langevin dynamics.
This could be another reason influence the accuracy of the modeling results.

For the parametric model, though it has reproduced many key features of the experimental results,
there are still improvements could be considered in future work.
One simplification used in the derivation of the model equations is that,
we have neglected the weakening of post-directional annealing for the last few incoming particles.
(This is because there is no incoming particles after the irradiation stops,
thus the number of damage cascades involved in post-directional annealing are overestimated for the last few incoming particles.)
This approximation may cause degraded accuracy of the calculated fluence dependence of EF.

In the modeling, we have attribute the delayed annealing of defects to the construction of the metastable complex.
The exact types of these complex are remain to be investigated.
Besides interstitial clusters, boron-interstitial clusters are also possible candidates.
The research of transient enhanced diffusion (TED) effects pointed out that
B cluster precursors, such as B$_m$I$_n$ ($n$ is large) are formed when Si interstitial supersaturation is high.
In the annealing process followed, the cluster emit Si interstitials and nucleate into electrically inactive stable complex,
such as B$_m$I.~\cite{Pelaz1997,Agarwal1997,Pelaz1999}

\section{Conclusion}
\label{sec:conclusion}

The experimentally observed enhanced low-flux sensitivity (ELFS) effects of neutron-induced displacement damage in bipolar devices have been explained and simulated in details.
The ELFS effects are attributed to
the suppression of Si-interstitials mediated inter-cascade interactions at low flux. 
At low flux, the Si interstitials emitted from one cascade have sufficient time
to dissipate before the subsequent cascades take place.
However, at high flux, 
many subsequent cascades have established and 
enhance the probability of capturing the interstitials.
The conflicts between the rapid annealing results of previously assumed diffusion dynamics limited model ($t_{anneal}<10^{0}s$) and
the slow annealing speeds observed in our experiments ($t_{anneal}\sim10^3s$) implies that
the cluster's nature of the defects has remarkable influence on the annealing processes.
The accordance of the characteristic time of ELFS effects and the damage annealing experiments
implies that the inter-defect mechanisms take actions through the annealing processes,
and the sensitive range of flux 
is limited by the rates of the annealing processes of defect clusters.
Based on these viewpoints,
modified numerical simulations not only reproduced the flux dependence of the defects formation,
but also gave a transition flux which matches the experimental configurations.

By summarizing the mechanisms, a parametric model is established.
The model has compact analytic form containing only a few parameters characterizing the systems.
The calculated results not only reproduce the flux dependence of the EF,
but also predict the fluence and temperature dependence of EF.
The results show that the flux dependence is remarkably suppressed for sufficient high fluence.
The reasons are the saturation of recombinations through pre-directional annealing processes which
overwhelm the post-directional annealing processes.
The results can explain the reason why ELFS effects have rarely been observed in previous experiments.

Our work should be helpful in analyzing the displacement damage of the irradiation vulnerable systems.
It shows that, when estimating the damage, not only the total fluence, but also the flux and temperature should be considered as well.
The observations may also provide some knowledge of controlling the defect concentrations through ion implantations by
tuning temperatures and flux rates.

\begin{acknowledgments}
The work was financially supported by 
the Science Challenge Project under Grant No. TZ2016003-1 and 
NSFC under Grant Nos. 11804313 and 11404300.
\end{acknowledgments}

\begin{appendix}

\section{TCAD simulation of GLPNP BJT}
\label{app:TCAD}

The configuration of the specially designed GLPNP BJT is shown in Fig.~\ref{fig:GLPNP_config}.
The emitter, base and collectors have the same length of $8~\mu m$ and the
intervals between them are equal to $12~\mu m$.
Both emitter and collectors are heavily $p+$ doped to approximate $10^{19}~cm^{-3}$ and 
the base is $n+$ doped to approximate $10^{18}~cm^{-3}$.
The buried n-type Si layer (BN) is $A_s$ doped to $4\times10^{15}~cm^{-3}$ and
the epitaxial n-layer (N-epi) has slighter doping concentration.
The gate oxides are produced by a layer of $7000\AA$ field oxide
plus a layer of $7500\AA$ TEOS.
Aluminum gates are deposited on top of each contacts and
transistors are separated by insulators and buried p layers.

\begin{figure}[h]
  \includegraphics[width=0.4\textwidth]{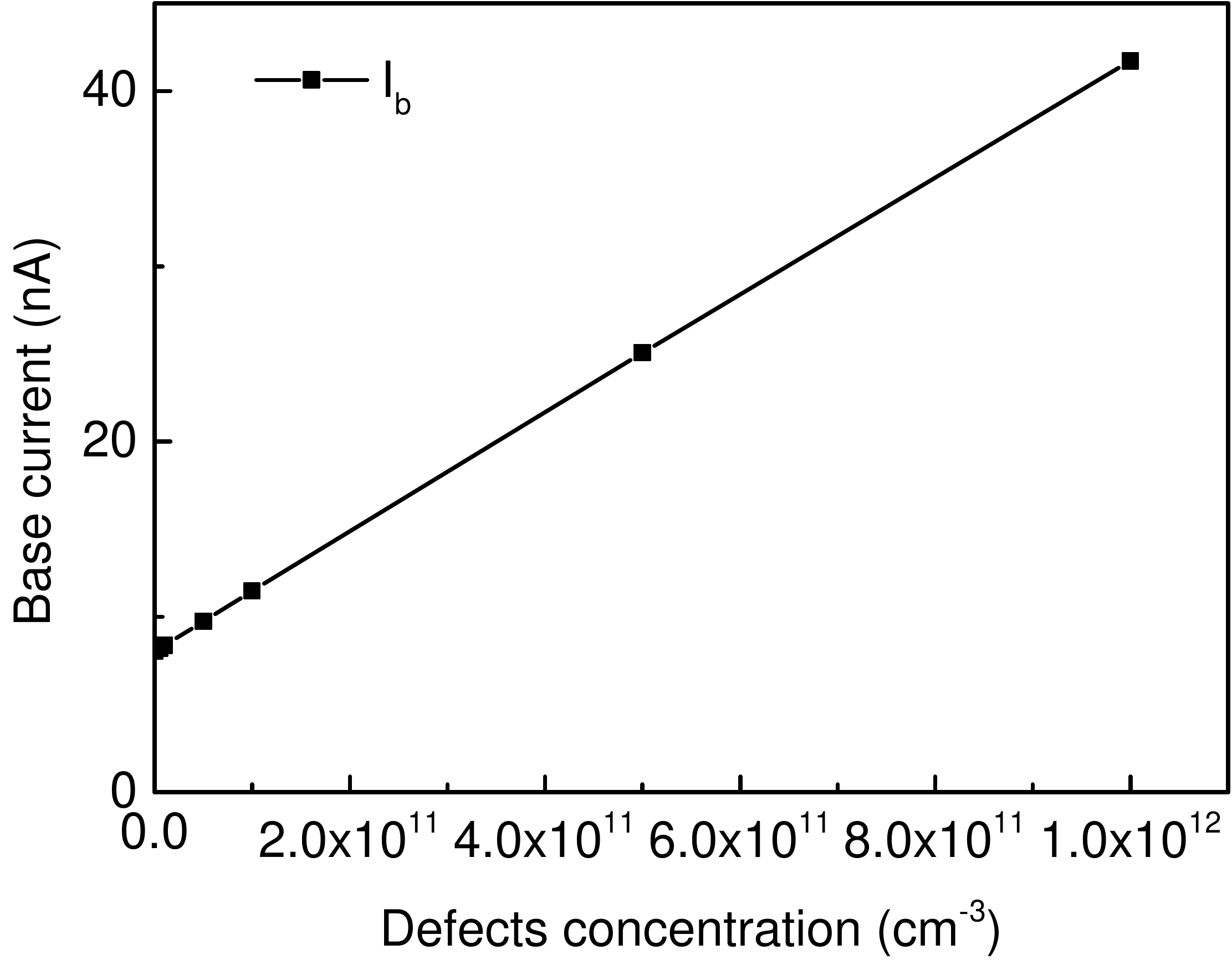}
  \caption{Simulated base current $I_b$ vs defects concentration.}
  \label{fig:GLPNP_Ib}
\end{figure}

The simulation is made by Santaurus TCAD and the results are plotted in Fig.~\ref{fig:GLPNP_Ib}.
The defects inserted in the simulations are assumed to have electronic parameters similar to $V_2$ defect.
The results show that the degradation of base current is a linear dependence on the concentration of the defects
of silicon bulk.
This means the observed flux effects of $\Delta I_B$ during irradiation are
the flux effects of defects generation in silicon region.

\section{The conversion between the irradiation flux and $\Delta t$ between subsequent impacts of incoming particles}
\label{app:fluxcalculation}

In the experiments, the irradiation rate is measured by flux while in simulations
the rate is measured by $\Delta t$ of impacts from incoming particles.
This section interprets simple approaches estimating $\Delta t$ from the known flux and vice verse.

Using our experimental setting as an example,
the neutron flux is between $5\times10^5$ and $5\times10^6~cm^{-2}~s^{-1}$ which results in
an average time intervals between incoming particles varying between $10^1$ and $10^3~s$ (for $1\mu m^{2}$ area).
In silicon, the $1~\text{MeV}$ neutron has a collision probability approximated to $0.18$ per $cm$ thickness
which enlarges $\Delta t$ between collisions of incoming neutrons with the lattice atoms to $10^5\sim10^7~s$
 (for $1\mu m^{3}$ volume).
The mean distance among defect trajectories (as shown in Fig.~\ref{fig:Impact_KMC})
for a fluence of $1\times10^{10}~cm^{-2}$ is $1-10\mu m$ for $1\mu m$ thickness of the sample.
Considering that the size of our samples and the sensitive region are approximately $10~\mu m$ in each dimension,
$\Delta t$ between adjacent collisions would be $10^2-10^4s$.

The simulated results show ELFS effects when $\Delta t$ is in range of $\sim10^{-5}-10^{-2}s$
(Fig.~\ref{fig:Sim(nodop)}),
base on the similar calculations described above,
the results corresponding to the transition flux, $\Phi_t$, of values of $\sim10^{11}\text{-}10^{14}~cm^{-2}s^{-1}$.

\section{Energy spectrum of PKAs for 1 MeV neutron bombardment}
\label{app:GENT4}

\begin{figure}[h]
  \includegraphics[width=0.45\textwidth]{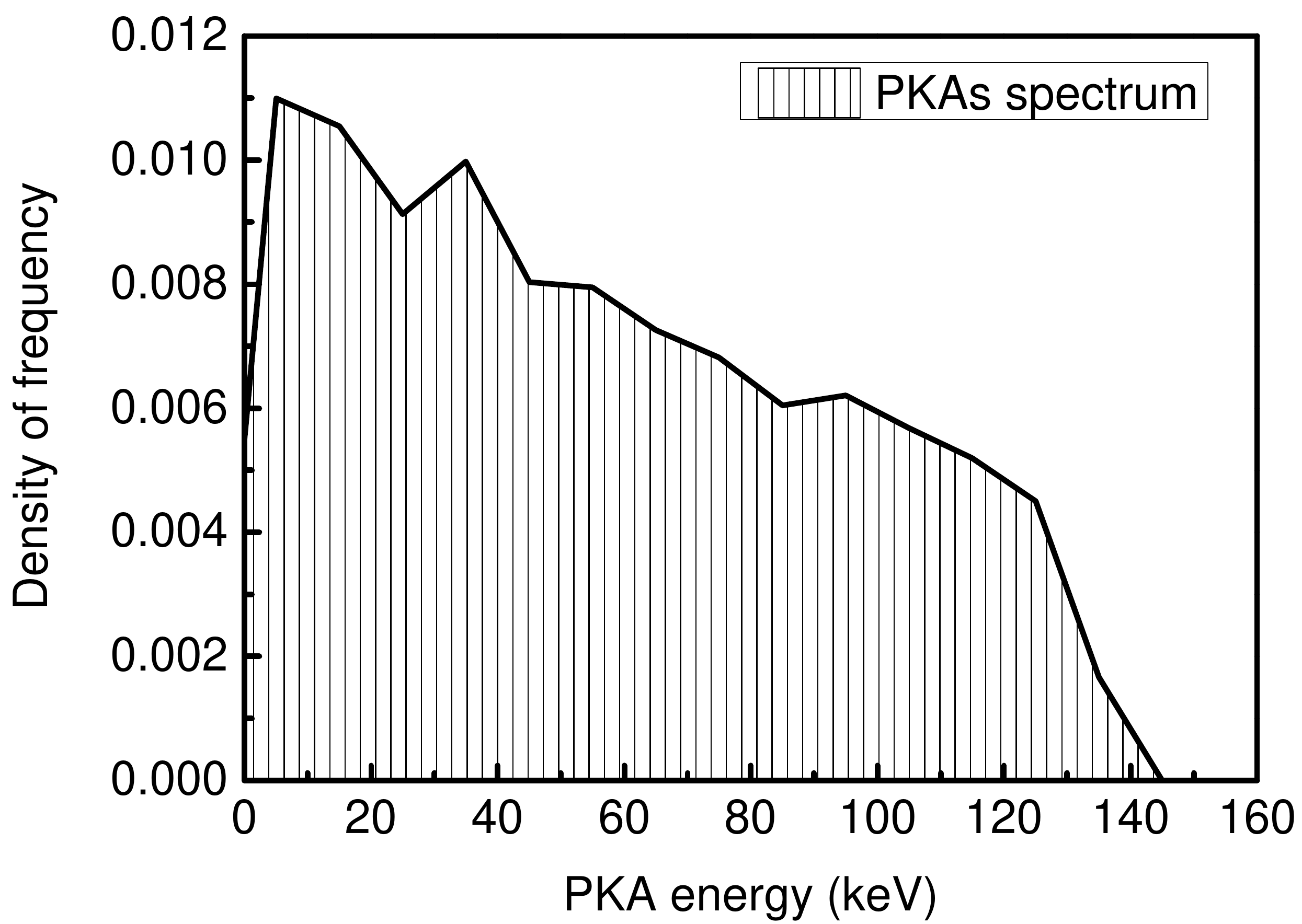}
  \caption{The spectrum of PKAs after $1MeV$ neutron bombardments.
The integrated area of the energy interval represents the probability of PKAs of energies within that interval.}
  \label{fig:PKAspectrum}
\end{figure}

The spectrum of knock-on atoms (PKAs) for 1MeV neutrons' impacts on silicon are calculated by GENT4.
In the simulations,
a total fluence of $10^{12}cm^{-2}$ is choosen to obtain sufficient times of impacts for statistics.
The resulted density of frequency of PKAs' energy are plotted in Fig.~\ref{fig:PKAspectrum}.
The results show that the PKAs stimulated by 1MeV neutrons have broad spectrum.
Most of them have energies exceeding tens of keVs which is supposed to have overcome the threshold of creating defect clusters
\cite{Wood1980}.

\section{The derivative of the parametric model equations}
\label{app:parametricmodel}

This section explains the derivative of the basic equations (Eqs.~\ref{eq:parametricmodel}-\ref{eq:parametricmodel_3}) of
parametric model in Chap.~\ref{sec:Parametricmodel}.

The mechanisms and assumptions of the model have been described in the main text
(Chap.~\ref{chap:mechanisms}, \& \ref{sec:Parametricmodel}).
The schematic illustrations of the model are shown in Fig.~\ref{fig:mechanism_interactions} \& \ref{fig:model_analytic}.
Defects are introduced as the results of damage cascades of incoming particles.
After each cascade,
defect clusters are formed containing interstitial and vacancy-related defects.
The annealing of the cluster is considered to happen through the emissions of the Si interstitials and
their recombinations with vacancy-related defects.
We define a parameter $\alpha$ as the ratio of the Si interstitials that can be emitted into the bulk
which limits the maximum of Si interstitials involved in the recombination processes.
The recombination processes take place through two kinds of reactions:
1) The pre-directional recombinations in which Si interstitials recombine with 
the defects created before the creations of themselves
(including the vacancy-related defects created within the same cascade of themselves).
2) The post-directional recombinations in which Si interstitials recombine with the defects created from subsequent damage cascades.

\subsection{Ratio parameter $\chi$ of recombinations}

To measure the ratios of the defects introduced by a single impact which
contribute to the pure increase of the whole defects in system,
we define a dimensionless parameter $\chi$.
$\chi$ has the following expression:
\begin{equation}\label{eq:chi}
\chi=(1-\alpha)+\alpha e^{-(n_a+n_{r,eff})/n_s}~,
\end{equation}
where $n_a$ is the number of existing defects,
$n_{r,eff}$ is the effective number of defects involved from subsequent damage cascades.
$n_s$ defines the characteristic number of defects to induce prominent recombinations.
The first term in Eq.~\ref{eq:chi} represents the portion of Si interstitials not involved in recombinations
(not emitted through annealing).
The second term calculates the number of Si interstitials survived from both
the pre- and post-directional recombinations.
The efficiency of the recombinations is assumed to be
exponentially proportional to $n_a$ and $n_{r,eff}$ scaled by $n_s$:
$1-e^{-(n_a+n_{r,eff})/n_s} ~$.

\subsection{Effective defects $n_{r,eff}$}

$n_{r,eff}$ is estimated by the integral:
\begin{equation}\label{eq:nr_int}
n_{r,eff}=\int^{t'_{max}}_{t'_{min}} \gamma(t)\kappa G^r(t')dt' ~,
\end{equation}
where $t$ is the intervals between the first generated defects and the defects introduced from the subsequent irradiations.
$\gamma(t)\kappa dt'$ calculates the number of defects introduced within the interval $dt'$ at time $t$.
As the annealing of a cluster proceeds,
with the increased time intervals between the formations of itself and the later generated defects,
the total number of its emitted interstitials that can encounter the later generated defects decrease.
$G^r(t')$ measures the efficiency of the newly generated defects at time $t'$ with respect to the defects generated
previously at $t'=0$,
which has the form:
\begin{equation}\label{eq:gr}
G^r(t')=e^{-t'/\tau} ~.
\end{equation}
Substituting Eq.~\ref{eq:gr} into Eq.~\ref{eq:nr_int}, it can be obtained:
\begin{equation}\label{eq:nr_int1}
n_{r,eff}=\gamma\kappa\tau\left( e^{-t'_{min}/\tau}-e^{-t'_{max}/\tau} \right) ~.
\end{equation}
For simplicity,
we define the upper limit of the integral $t'_{max}$ equals to infinity.
The value of the lower limit $t'_{min}$ depends on the rates (or time intervals) of the impacts from incoming particles,
hence is a function of the flux of the radiation.
$t'_{min}$ is determined through the function:
\begin{equation}\label{eq:tmin}
t'_{min}=1/(\gamma\kappa) ~.
\end{equation}
Combining Eqs.~\ref{eq:nr_int1} and \ref{eq:tmin}, $n_{r,eff}$ is obtained:
\begin{equation}\label{eq:nr}
n_{r,eff}=\gamma\kappa\tau e^{-1/(\gamma\kappa\tau)} ~.
\end{equation}

\subsection{Defects survived from recombinations after irradiation $n^{rem}$}

The number of the defects survived from the recombinations satisfies the following differential equation:
\begin{equation}\label{eq:dnr}
\frac{dn^{rem}(t)}{dt}=\chi(t)\gamma ~.
\end{equation}
Using Eq.~\ref{eq:chi}, \ref{eq:nr} and the approximation $n_{rem}\approx n_a$,
Eq.~\ref{eq:dnr} is solved:
\begin{widetext}
\begin{equation}\label{eq:nremain}
n^{remnant}=n^{s}\text{ln}\left(\frac{1}{-1+\alpha}\left(e^{-n^{r,eff}/n^{s}}\alpha+e^{-n^{tot}(-1+\alpha)/n^{s}}\left(-1+\alpha-e^{-n^{r,eff}/n^{s}}\alpha\right)\right)\right) ~,
\end{equation}
\end{widetext}
where the boundary condition $n_a(t=0)=0$ is used.
Eq.~\ref{eq:nremain} is exact Eq.~\ref{eq:parametricmodel} in the main text.

\end{appendix}

\end{document}